\pdfoutput=1

\documentclass[11pt,twoside,a4paper,cmspaper,final,collab]{cms-tdr}
\def\svnVersion{273060}\def\cmsCernNo{2013-037}\def\cmsCernDate{\today}\def\cmsMessage{Published in Physics Letters B as \href{http://dx.doi.org/10.1016/j.physletb.2014.12.003}{\doi{10.1016/j.physletb.2014.12.003.}}}

\begin{document}\cmsNoteHeader{SMP-12-023}

\hyphenation{had-ron-i-za-tion}
\hyphenation{cal-or-i-me-ter}
\hyphenation{de-vices}

\RCS$Revision: 268573 $
\RCS$HeadURL: svn+ssh://svn.cern.ch/reps/tdr2/papers/SMP-12-023/trunk/SMP-12-023.tex $
\RCS$Id: SMP-12-023.tex 268573 2014-11-20 18:57:56Z alverson $
\newlength\cmsFigWidth
\ifthenelse{\boolean{cms@external}}{\setlength\cmsFigWidth{0.90\columnwidth}}{\setlength\cmsFigWidth{0.48\textwidth}}
\newlength\cmsFigWidthTwo
\ifthenelse{\boolean{cms@external}}{\setlength\cmsFigWidthTwo{0.45\textwidth}}{\setlength\cmsFigWidthTwo{0.45\textwidth}}
\newlength\cmsFigWidthTall
\ifthenelse{\boolean{cms@external}}{\setlength\cmsFigWidthTall{0.44\textwidth}}{\setlength\cmsFigWidthTall{0.45\textwidth}}
\ifthenelse{\boolean{cms@external}}{\providecommand{\cmsLeft}{top}}{\providecommand{\cmsLeft}{left}}
\ifthenelse{\boolean{cms@external}}{\providecommand{\cmsRight}{bottom}}{\providecommand{\cmsRight}{right}}
\providecommand{\BLACKHAT} {\textsc{BlackHat}\xspace}
\providecommand{\FEWZ}{\textsc{fewz}\xspace}
\providecommand{\MT}{\ensuremath{M_\mathrm{T}}\xspace}
\cmsNoteHeader{SMP-12-023}
\title{Differential cross section measurements for the production of a W boson in association with jets in proton-proton collisions at $\sqrt{s} = 7$\TeV}

\date{\today}

\abstract{
Measurements are reported of differential cross sections for the production of a
W boson, which decays into a muon and a neutrino, in association with jets, as a function of several variables,
including the transverse momenta (\PT) and pseudorapidities of the four
leading jets, the scalar sum of jet transverse momenta (\HT), and the
difference in azimuthal angle between the directions of each jet and the muon.
The data sample of pp collisions at a centre-of-mass energy of 7\TeV was collected
with the CMS detector at the LHC and corresponds to an integrated luminosity of 5.0\fbinv.
The measured cross sections are compared to predictions from Monte Carlo generators,
\MADGRAPH{}+\PYTHIA and \SHERPA, and to  next-to-leading-order calculations from
\BLACKHAT{}+\SHERPA.
The differential cross sections are found to be in agreement with the predictions, apart from the \pt distributions of the leading jets at high \pt values, the distributions of the \HT at high-\HT and low jet multiplicity, and the distribution of the difference in azimuthal angle between the leading jet and the muon at low values.}

\hypersetup{%
pdfauthor={CMS Collaboration},%
pdftitle={Differential cross section measurements for the production of a W boson in association with jets in proton-proton collisions at sqrt(s) = 7 TeV},%
pdfsubject={CMS},%
pdfkeywords={CMS, physics, W+Jets, vector boson, jets, electroweak, standard model}}

\maketitle

\section{Introduction}
\label{introduction}

This letter reports measurements of fiducial cross sections for $\PW$ boson
production in association with jets at the LHC. Measurements of the production of vector bosons in association with
jets are fundamental tests of perturbative quantum chromodynamics (pQCD).
The $\PW$+jets processes also provide the main background to other, much rarer, standard model (SM) processes,
such as $\ttbar$~\cite{cmsttbar} and single top-quark
production~\cite{cmssingletop}, and to Higgs boson production and a variety of physics
processes beyond the SM. Searches for phenomena beyond the SM are
often limited by the uncertainty in the theoretical cross sections for
$\PW$ (and $\Z$) + jets processes at high momentum scales and large jet
multiplicities. Therefore, it is crucial to perform precision
measurements of $\PW$+jets production at the LHC.

Leptonic decay modes
of the vector boson are often used in the measurement of SM processes
and in searches for new physics, because they provide clean signatures
with relatively low background.
This letter focuses on the production of a $\PW$ boson decaying into
a muon and a neutrino, as part of a final-state topology
characterised by one high-transverse-momentum (\pt) isolated muon, significant missing
transverse energy ($\MET$), and one or more jets.
The cross sections are measured as a function of the inclusive and exclusive jet
multiplicities for up to six jets.
Differential cross sections are measured for different inclusive jet multiplicities as a function of
the transverse momentum and the pseudorapidity ($\eta$) of the jets,
where $\eta = -\ln [ \tan(\theta/2)]$, and $\theta$ is the polar angle measured with
respect to the anticlockwise beam direction.
The cross sections are also measured as a function of the difference in azimuthal
angle between the direction of each jet and that of the
muon, and of \HT, which is defined as the scalar sum of the \pt of all jets with $\pt>30\GeV$ and $\abs{\eta}<2.4$. It is important to study the distribution of the jet \pt and the observable \HT because they are sensitive to higher order corrections, and are often used to discriminate against background in searches for signatures of physics beyond the SM. Additionally, \HT is often used
to set the scale of the hard scattering process in theoretical calculations.
Finally, the $\eta$ distributions of jets and the azimuthal separations between the jets and the muon are also important, because they are sensitive to the modelling of parton emission.

The measurements presented in this letter use proton-proton (pp) collision data at a centre-of-mass energy of $\sqrt{s}=7$\TeV
recorded with the CMS detector at the LHC in 2011 and correspond to an
integrated luminosity of {$5.0 \pm 0.1\fbinv$}~\cite{lumi_new}.
These measurements cover high jet multiplicities and higher jet $\PT$ than earlier publications because the centre-of-mass energy and the integrated luminosity are higher.
Previous studies of leptonic decay modes of the W boson in association with jets at the LHC
have measured the cross sections and cross section ratios for $\PW$ boson production in association
with jets in pp collisions with an integrated luminosity of $36\pbinv$ at $\sqrt{s}=7$\TeV with the ATLAS~\cite{ATLAS36pbWJets} and CMS~\cite{CMS36pbWJets} detectors. Measurements have also been made with $\Pp\Pap$ collisions with the D0 detector~\cite{Abazov2011200,PhysRevD.88.092001} at the Tevatron collider for integrated luminosities up to 4.2\fbinv, as well as with the CDF detector~\cite{CDFwplusjets} for an integrated luminosity of 320\pbinv. Recent measurements have been made with the ATLAS detector with a centre-of-mass energy of $7$\TeV and an integrated luminosity of $4.6\fbinv$~\cite{AtlasWJets4p6pb}.

In order to perform a differential measurement
of the $\PW$+jets cross section, a high-purity sample of
$\PW\to \mu\nu$ events is selected and the kinematic
distributions are corrected to the particle level by means of
regularised unfolding~\cite{SVDUnfold}. This procedure corrects a measured observable
for the effects of detector response, finite experimental resolutions,
acceptance, and efficiencies, and therefore allows for direct comparison with
theoretical predictions. The measured differential cross sections are compared to the predictions of
generators such as  \MADGRAPH 5.1.1~\cite{MG5} interfaced with \PYTHIA 6.426~\cite{PYTHIA}, \SHERPA 1.4.0~\cite{Gleisberg:2008fv,Schumann:2007mg,Gleisberg:2008ta,Hoeche:2009rj}, and \BLACKHAT~\cite{BLACKHAT,PhysRevLett.106.092001}, interfaced to \SHERPA.
The \BLACKHAT{}+\SHERPA samples~\cite{Bern:2013zja} provide parton-level predictions of $\PW+n$ ($n=1$--5)
 jets at next-to-leading order (NLO),  while the \MADGRAPH{}+\PYTHIA
and \SHERPA samples provide
tree-level calculations followed by hadronisation to produce the final states.

The letter proceeds as follows: Section~\ref{cms} presents the CMS
detector. Section~\ref{samples} describes the Monte Carlo
(MC) event generators, as well as the data
samples used for the analysis. The identification criteria for the
final-state objects (leptons and jets) and the selection of the
$\PW\to \mu\nu +$ jets events are presented in
Section~\ref{eventselection}.
Section~\ref{background} describes the modelling of instrumental backgrounds and irreducible physics backgrounds.
The procedure used for unfolding is detailed in Section
~\ref{unfolding}, and Section~\ref{systematics} describes the
systematic uncertainties.  Finally, the unfolded distributions are presented in Section~\ref{results} and compared to theoretical predictions, and Section~\ref{conclusion} summarises the results.
\section{The CMS detector}
\label{cms}
The CMS detector, presented in detail
elsewhere~\cite{CMS}, can be described with a cylindrical coordinate system with the +$z$
axis directed along the anticlockwise beam axis. The detector consists of an
inner tracking system and calorimeters (electromagnetic, ECAL, and hadron, HCAL)
surrounded by a 3.8\unit{T} solenoid. The inner tracking system
consists of a silicon pixel and strip tracker, providing the required
granularity and precision for the reconstruction of vertices of
charged particles in the range $0 \leq \phi < 2\pi$ in azimuth and
$\abs{\eta}<2.5$. The crystal ECAL and the brass/scintillator
sampling HCAL are used to measure the energies of
photons, electrons, and hadrons within $\abs{\eta}<3.0$.
The HCAL, when combined with the ECAL, measures jets with a resolution $\Delta E/E \approx 100\% / \sqrt{E\,[\GeVns]} \oplus 5\%$~\cite{Chatrchyan:2013dga}.
The three muon
systems surrounding the solenoid cover a region $\abs{\eta}<2.4$ and are
composed of drift tubes in the barrel region $(\abs{\eta}<1.2)$,
cathode strip chambers in the endcaps $(0.9<\abs{\eta}<2.4)$, and
resistive-plate chambers in both the barrel region and the endcaps
$(\abs{\eta}<1.6)$.
Events are recorded based on a trigger decision using
information from the CMS detector subsystems.
The first level (L1) of the CMS trigger system, composed of custom hardware processors, uses information from the calorimeters and muon detectors to select the most interesting events in a fixed time interval of less than 4\mus. The high-level trigger (HLT) processor further decreases the event rate from 100\unit{kHz} at L1 to roughly 300\unit{Hz}.

\section{Data and simulation samples}
\label{samples}

Events are retained if they pass a trigger requiring one isolated
muon with $\pt>24$\GeV and $\abs{\eta}<2.1$.
Signal and background simulated samples are produced and fully reconstructed
using a simulation of the CMS detector based on {\GEANTfour}~\cite{GEANT4}, and simulated events are required to pass an emulation of
the trigger requirements applied to the data.
These simulations include multiple collisions
in a single bunch crossing (pileup). To model the effect of pileup, minimum bias events generated in \PYTHIA are added to the simulated events,
with the number of pileup events selected to match the pileup multiplicity distribution observed in data.

A $\PW\to \ell \nu$+jets signal sample is generated with \MADGRAPH5.1.1
 and is used to determine the detector
response in the unfolding procedure described in Section~\ref{unfolding}.
Parton showering and hadronisation of the \MADGRAPH samples
are performed with \PYTHIA 6.424
 using the Z2 tune~\cite{tunez2}.
The detector response is also determined using a different
$\PW$+jets event sample generated with \SHERPA 1.3.0~\cite{Gleisberg:2008fv,Schumann:2007mg,Gleisberg:2008ta,Hoeche:2009rj},
and is used in the evaluation of systematic uncertainties due to the unfolding of the data.

The main sources of background are the production
of $\ttbar$, single top-quark, $\cPZ/\gamma^*$+jets, dibosons ($\cPZ\cPZ/\PW\cPZ/\PW\PW$) +
jets, and multijet production.
With the exception of multijet production, all backgrounds
are estimated from simulation. The
simulated samples of $\ttbar$ and $\cPZ/\gamma^*$+jets are generated with \MADGRAPH 5.1.1;
single top-quark samples ($s$-, $t$-, and t$\PW$- channels) are generated with {\POWHEG} version 1.0
~\cite{Nason:2004rx,Frixione:2007vw,Alioli:2010xd,Alioli:2009je}; $\text{VV}$ samples, where $\text{V}$ represents either a $\PW$ boson or
a $\cPZ$ boson, are generated with \PYTHIA version 6.424
 using the Z2 tune~\cite{tunez2}. Parton showering and hadronisation of the \MADGRAPH and \POWHEG samples
are performed with \PYTHIA 6.424.
The simulations with \MADGRAPH and \PYTHIA use the CTEQ6L1 parton distribution functions
(PDF)~\cite{CTEQ}. The simulation with \SHERPA uses the CTEQ6.6m PDF, and the simulations with \POWHEG use the CTEQ6m PDF.

The $\PW$+jets and $\cPZ/\gamma^*$+jets samples are normalised to next-to-next-to-leading order (NNLO) inclusive
cross sections calculated with \FEWZ~\cite{FEWZ}. Single top-quark and
$\text{VV}$ samples are normalised to NLO inclusive cross sections calculated
with \MCFM~\cite{ mcfm_t, mcfm_Wt, mcfm_tch,mcfm_diboson}.
The $\ttbar$ contribution is normalised to the NNLO + next-to-next-leading logarithm (NNLL) predicted cross section from Ref.~\cite{Czakon:2013goa}.

\section{Object identification and event selection}
\label{eventselection}

Muon candidates are reconstructed as tracks in the muon system that are matched
to tracks reconstructed in the inner tracking
system~\cite{CMS-PAPERS-MUO-10-004}. Muon candidates are required to have $\pt >25\GeV$, and to
be reconstructed within the fiducial volume used for the high-level trigger muon selection, \ie within
$\abs{\eta} < 2.1$. This ensures that the offline event selection
requirements are as stringent as the trigger.
In addition, an isolation requirement is applied to the muon candidates by demanding
that the relative isolation is less than 0.15, where the relative isolation
is defined as the sum of the transverse energy deposited in the calorimeters (ECAL and HCAL)
and of the \pt of charged particles measured with the tracker in a cone of
 {$\Delta R = \sqrt{\smash[b]{(\Delta \phi )^2 + (\Delta
\eta) ^2}} = 0.3$} around the muon candidate track (excluding this track),
divided by the muon candidate \pt. To ensure a precise measurement of the
transverse impact parameter of the muon track relative to the interaction point,
only muon candidates with tracks containing more than 10 hits in the
silicon tracker and at least one hit in the pixel detector are
considered. To reject muons from cosmic rays, the transverse impact
parameter of the muon candidate with respect to the primary vertex is required to be less
than 2\unit{mm}.

Jets
are reconstructed using the CMS particle-flow
algorithm~\cite{CMS-PAS-PFT-09-001, CMS-PAS-PFT-10-001},
using
the anti-\kt~\cite{antikt,fastjetmanual} algorithm with a distance parameter
of 0.5. The jet energy is calibrated using the \pt balance of
dijet and $\gamma+$jet events~\cite{JetCorr} to account for the following effects:
nonuniformity and nonlinearity of the ECAL and HCAL energy response to
neutral hadrons, the presence of extra particles
from pileup interactions, the thresholds used in jet constituent selection,
 reconstruction inefficiencies, and possible biases introduced by the
clustering algorithm.
Only jets with $\pt >30\GeV$, $\abs{\eta} < 2.4$, and a
spatial separation of $\Delta R > 0.5$ from the muon are considered. To reduce the
contamination from pileup jets, jets are required to be associated to the same primary vertex as the muon.
The vertex associated to each jet is the one that has the largest number of \pt-weighted tracks in
common with the jet. The contamination from pileup jets is estimated with the signal simulation, with pileup events
simulated with \PYTHIA, and found to be less than 1\%.

The missing momentum vector, $p_\mathrm{T}^\text{miss}$, is defined as the negative of the vectorial sum of the transverse momenta of the particles reconstructed with the particle-flow algorithm, and the $\MET$ is defined as the magnitude of the $p_\mathrm{T}^\text{miss}$ vector. The measurement of the $\MET$ in simulation is sensitive to the modelling of the calorimeter response and resolution and to the description of the underlying event.
To account for these effects, the $\MET$ in $\PW$+jets simulation is corrected for the differences in the detector response between data and simulation, using a method detailed in Ref.~\cite{CMS-PAPERS-EWK-10-005}.
A recoil energy correction is applied to the $\PW$+jets simulation on an event-by-event basis,
using a sample of $\Z \to \mu \mu$ events in data and simulation. The transverse recoil vector, defined as
the negative vector sum of the missing transverse energy and the transverse momenta
of the lepton(s), is divided into components
parallel and perpendicular to the boson direction. The mean and the width of the transverse
recoil vector components are parameterised as a function of the \Z boson \pt in data and simulation.
The ratio of the data and simulation parameterisations is used to adjust the transverse
recoil vector components in each simulated event, and a new $\MET$ is computed using the
corrected recoil components.

Events
are required to contain exactly one muon satisfying the conditions
described above and one or more jets with $\pt >30\GeV$.
Events are required to have $\MT >50\GeV$,
 where \MT, the
transverse mass of the muon and missing transverse energy, is defined as
$\MT \equiv \sqrt{\smash[b]{2 \pt^{\mu}   \MET \left( 1 - \cos{\Delta \phi} \right) }}$,
 where $\pt^{\mu}$ is the muon \pt and $\Delta \phi$ is the difference in azimuthal
 angle between the muon momentum direction and the $p_\mathrm{T}^\text{miss}$ vector.

\section{Estimation of the backgrounds and selection efficiencies}
\label{background}

All background sources except for the multijet production are modelled with simulation.
The simulated event samples
are corrected for differences between data and simulation in muon identification
efficiencies and event trigger efficiency.
A ``tag-and-probe" method~\cite{CMS-PAPERS-MUO-10-004} is used to determine
the differences between simulation and data for the efficiency of the trigger and for the muon
identification and isolation criteria.
This method uses $\Z \to \mu \mu$
events from both data and simulated samples where the ``tag" muon is required to pass the identification and
isolation criteria. The efficiency measurements use the ``probe" muon, which is required to
pass minimal quality criteria. Trigger efficiency corrections are determined as a function of the
muon $\eta$, and are in general less than 5\%. Muon isolation and identification efficiency corrections
are determined as a function of the muon \pt and $\eta$, and are generally less than
2\%. Corrections to the simulation are applied on an event-by-event basis in the form of
event weights.

\begin{figure*}[htb]
\centering
\includegraphics[width=.48\textwidth]{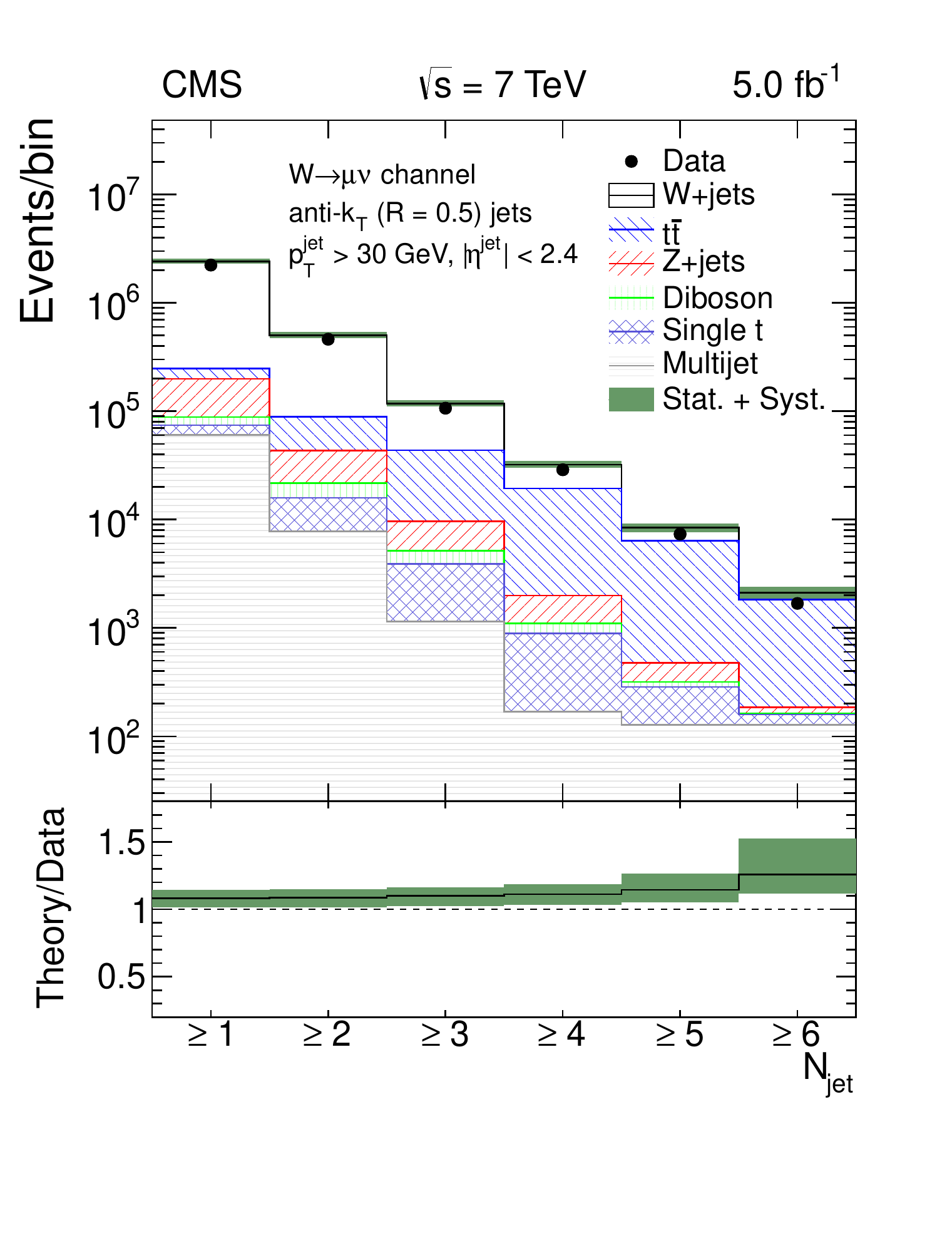}
\includegraphics[width=.48\textwidth]{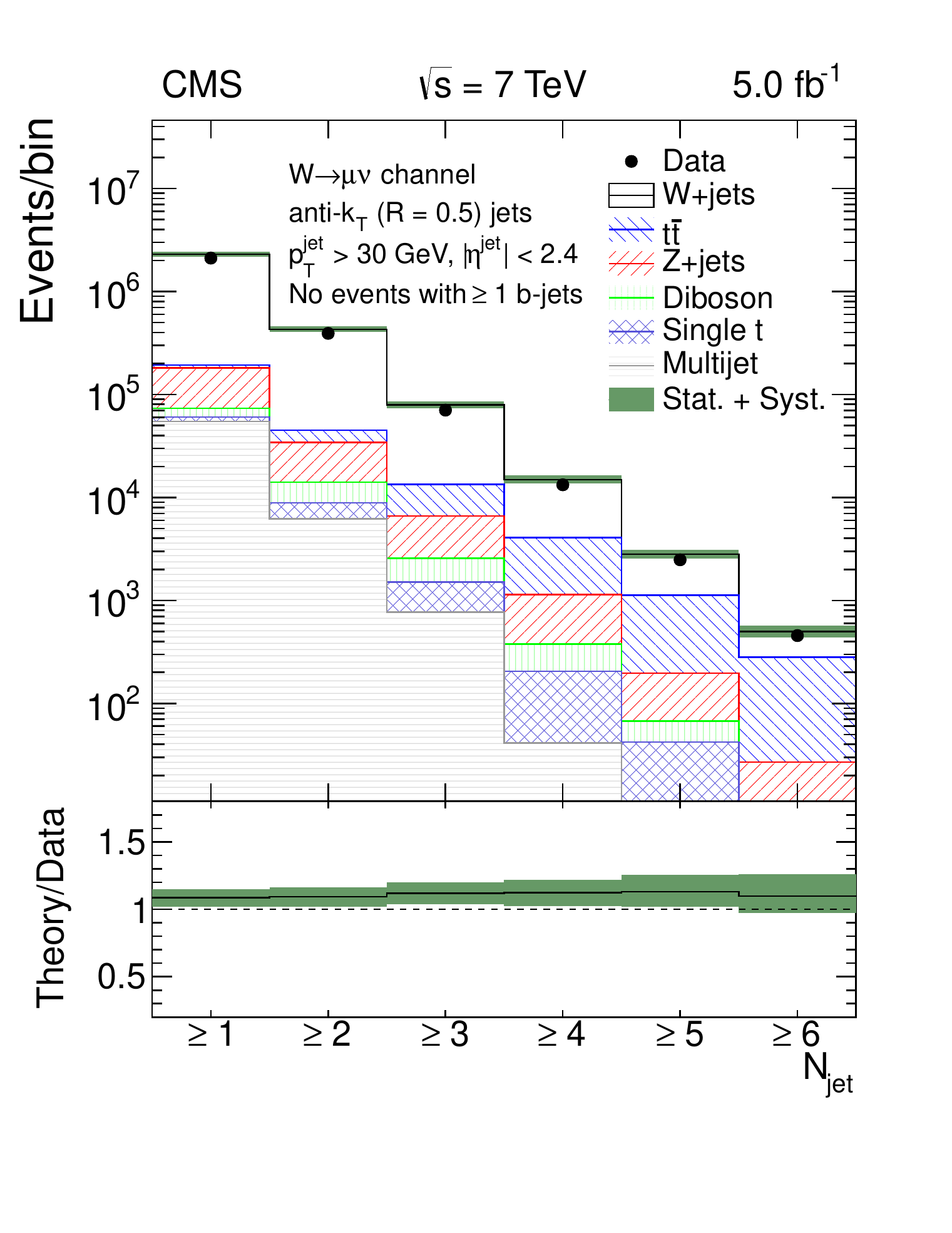}
\caption{The jet multiplicity in data and simulation before (left)  and after (right) the b-jet veto. The $\PW$+jets contribution is modelled with {\sc MadGraph} 5.1.1+{\sc Pythia} 6.424. The solid band indicates the total statistical and systematic uncertainty in the $\PW$+jets signal and background predictions, as detailed in Section~\ref{systematics}. This includes uncertainties in the jet energy scale and resolution, the muon momentum scale and resolution, the pileup modelling, the b-tagging correction factors, the normalisations of the simulations, and the efficiencies of reconstruction, identification, and trigger acceptance.  A substantial reduction in the expected $\ttbar$ background is observed in the right plot.}
\label{fig:BtagOnOff}

\end{figure*}

The dominant background to $\PW$+jets production is $\ttbar$ production, which has a larger total contribution than that of
the $\PW$+jets signal in events with four or more jets. In order to reduce the
level of $\ttbar$ contamination, a veto is applied to events with one or more
b-tagged jets. Heavy-flavour tagging is based on a tag
algorithm~\cite{btag} that exploits the long lifetime of
b-quark hadrons. This algorithm calculates the signed impact parameter significance of
all tracks in the jet that satisfy high-quality reconstruction and purity criteria, and orders the tracks by decreasing
significance. The impact parameter significance is defined as the ratio of the impact
parameter to its estimated uncertainty.
For jets with two or more significant tracks,
a high-efficiency b-quark discriminator is defined as
the significance of the second most significant track. The size of the $\ttbar$ background
is illustrated in Fig.~\ref{fig:BtagOnOff}, before and after the implementation of the
b-jet veto, using the event selection described in Section~\ref{eventselection}. The expected
contributions for the different processes in Fig.~\ref{fig:BtagOnOff} are shown as a function
of the jet multiplicity, along with the observed data.
 Differences in the tagging and mistagging rates between data and simulation
 are measured as a function of the jet \pt
 in multijet and $\ttbar$ events~\cite{btag}, and are used
 to correct the tagging rates of the jets in simulation.
For jet multiplicities of 1 to 6, the
b-jet veto eliminates 44--84\% of the predicted $\ttbar$
background, while eliminating 3--26\% of the predicted $\PW$+jets
signal. The resulting increase in the signal purity allows for reductions in the total uncertainty in the measured cross sections of 6--43\% for jet multiplicities of 4--6.

The shape and normalisations of the  $\Z/\gamma^*$+jets and $\ttbar$
predictions are cross-checked in selected data samples. The Z+jets background is compared to data in
a Z-boson dominated data sample that requires two well-identified, isolated muons. The $\ttbar$
background is compared to data in a control region requiring at least two b-tagged
jets. Background estimations from simulation and from data control samples agree within the uncertainties described in Section~\ref{systematics}.

The  multijet background is estimated using a data control sample
with an inverted muon isolation requirement. In the control sample, the muon misidentification rate for multijet processes is estimated in the multijet-enriched sideband region with $\MT<50$~GeV, and the shape template of the multijet distribution is determined in the region with $\MT>50$~GeV.
Contributions from other processes to the  multijet control region
are subtracted, including the dominant contribution from $\PW$+jets.
In order to improve the estimation of $\PW$+jets in the  multijet control region,
the $\PW$+jets contribution is first normalised to data in the $\MT >50\GeV$
region with the muon isolation condition applied.
The  multijet shape template is then rescaled according to the muon
misidentification rate.
For exclusive jet multiplicities of 1--4, the purity of the multijet-enriched inverted-isolation sideband region is 99.7--98.1\%, and the purity of the W+jets contribution to the signal region is 92--76\%.
The multijet estimate corresponds to 32.7--1.9\% of the total background estimate, or  2.6--0.3\% of the total SM prediction.

\section{The unfolding procedure}
\label{unfolding}

For the measurement of cross sections,
the particle level is defined by a $\PW$ boson, which decays
into a muon and a muon neutrino, produced in
association with one or more jets.
Kinematic thresholds on the particle-level muon, \MT, and jets are identical to those
applied to the reconstructed objects.
Specifically, the particle-level selection includes the requirement of
exactly one muon with $\pt>25\GeV$ and $\abs{\eta}<2.1$, and
$\MT >50\GeV$. The particle-level $\MET$ is defined as the negative
of the vectorial sum of the transverse momenta of all visible final state particles. To account for final-state radiation, the momenta of all photons in a cone of $\Delta R
< 0.1$ around the muon are added to that of the muon.   Jets are clustered using the  anti-$k\rm _T$~\cite{antikt} algorithm
with a distance parameter of $0.5$. Clustering is performed
using all particles after decay and fragmentation,
excluding neutrinos and the muon from the $\PW$ boson decay. Additionally,
jets are required to have $\pt >30\GeV$ and $\abs{\eta}<2.4$, and to be
separated from the muon by $\Delta R > 0.5$.

The reconstructed distributions are corrected to the particle level
with the method of regularised singular
value decomposition (SVD)~\cite{SVDUnfold} unfolding, using the
\textsc{RooUnfold} toolkit~\cite{RooUnfold}.
For each distribution, the total background, including the multijet estimate from data and all simulated processes except the $\PW$ boson signal,
is subtracted from the data before unfolding.
A response matrix, defining the migration probability between the
particle-level and reconstructed quantities, as well as the
overall reconstruction efficiency,  is computed using $\PW$+jets events simulated with
\MADGRAPH{}+\PYTHIA.
For a given particle-level quantity $Q$ with a corresponding reconstructed quantity $Q^{'}$, the migration probability from an interval $a<Q<b$ to an interval $c<Q^{'}<d$  is defined as the fraction of events with $a<Q<b$ that have $c<Q^{'}<d$.
The unfolding of the jet multiplicity is performed with a response defined by the number of particle-level jets versus the number of reconstructed jets. For particle-level jet multiplicities of 1 to 6, 4 to 51\% of simulated events exhibit migration to different values of reconstructed jet multiplicity.
The unfolding of the kinematic distributions of the $n$th jet  is performed with a response defined by the kinematic quantity of the $n$th-highest-\pt particle-level jet versus that of the $n$th-highest-\pt reconstructed jet. To achieve a full migration from the selection of reconstructed events to the particle-level phase space, no matching between reconstructed and particle-level jets is applied.
The contribution from the $\PW\to\tau\nu$ process with a muon in the final state is estimated to be at the 1\% level, and
is not considered as part of the signal definition at the particle level.

The b-jet veto is treated as an overall event selection condition.
Events failing this condition are treated as nonreconstructed in the
unfolding response, so that the cross section
obtained after unfolding is valid for $\PW$ boson decays with associated jets of any flavour.

\section{Systematic uncertainties}
\label{systematics}

The sources of systematic uncertainties considered in this analysis
are described below. The entirety of the unfolding procedure is
repeated for each systematic variation, and the unfolded data results
with these variations are compared with the central (unvaried) results
to extract the uncertainties in the unfolded data
distributions.

In most distributions, the dominant sources of systematic uncertainty include the jet energy scale and resolution uncertainties, which affect the shape of all reconstructed distributions as well as the overall event acceptance. The jet energy scale uncertainties are estimated by assigning a $\pt$- and $\eta$-dependent uncertainty in jet energy corrections as discussed in Ref.~\cite{JetCorr}, and by varying the jet $\pt$ by the magnitude of the uncertainty. The uncertainties in jet energy resolution are assessed by increasing the \pt difference between the reconstructed and particle-level jets by an $\eta$-dependent value~\cite{JetCorr}.
The jet energy uncertainties are determined by varying the \pt of the jets in data rather than in simulation.

Muon momentum scale and resolution uncertainties also introduce uncertainties in the overall event acceptance.
A muon momentum scale uncertainty of 0.2\% and a muon momentum resolution uncertainty of 0.6\% are assumed~\cite{CMS-PAPERS-MUO-10-004}. The effects of these uncertainties are assessed by directly varying the momentum scale and randomly fluctuating the muon momentum in the simulation.

Variations for uncertainties in the energy and momentum scales and resolutions affect the size and shape of the background distribution to be subtracted from the data distribution, as well as the acceptance of $\PW$+jets simulated events, which define the response matrix used for unfolding.
The variations are also propagated to the measurement of $\MET$, which affects the acceptance of the $\MT>50\GeV$ requirement.

Another important source of systematic uncertainty is
the choice of the generator used in the unfolding procedure. The
size of this uncertainty
is assessed by repeating the unfolding procedure with a response trained on a
separate simulated sample generated with
\SHERPA 1.3.0. The absolute value of the difference between the data unfolded with a response matrix
trained on \SHERPA and with a response matrix trained on \MADGRAPH{}+\PYTHIA
is treated as a symmetric uncertainty in the measurement.

Other minor sources of systematic uncertainty include the uncertainties in the background
normalisation, the b-tagging efficiency, the modelling of the Wb contribution in the signal simulation, integrated luminosity, the  pileup modelling, the trigger and
object identification efficiencies, and the finite number of simulated events used to construct the response matrix.
Background normalisation uncertainties are determined
by varying the cross sections of the backgrounds within their theoretical uncertainties.
For the $\Z$+jets process, a normalisation uncertainty of 4.3\% is calculated as the sum in quadrature of the factorisation/renormalisation scale and PDF uncertainties calculated in \FEWZ~\cite{FEWZ}. For the diboson and single top-quark processes, uncertainties are calculated with \MCFM~\cite{ mcfm_t, mcfm_Wt, mcfm_tch,mcfm_diboson} to be 4\% and 6\%, respectively.
The uncertainty in the $\ttbar$ modelling is assessed by taking the difference between data and simulation in a control region with two or more b-tagged jets, and is estimated to be 5 to 12\% for jet multiplicities of 2 to 6.
The estimate of the multijet background
has an uncertainty based on the limited number of events in the multijet sample and in the control regions where the multijet sample normalisation is calculated, and other systematic variations affecting the backgrounds in the multijet control regions introduce variations in the multijet normalization and template shape.
For the b-tagging algorithm used to veto events containing b jets,
uncertainties in the data/simulation ratio of the b-tagging efficiencies are applied.
For jets with $\pt >30\GeV$,
these uncertainties range from 3.1 to 10.5\%.
An additional uncertainty is ascribed to the normalisation of the Wb content in the simulation by examining the agreement between data and simulation as a function of jet multiplicity in a control region defined by requiring exactly one b-tagged jet. An increase in the normalisation of the Wb process of 120\% is considered, yielding an uncertainty in the measurement of 0.5 to 11\% for jet multiplicities of 1 to 6.  The uncertainty in the integrated luminosity is
2.2\%~\cite{lumi_new}. An uncertainty in the modelling of pileup in simulation is determined by varying the number of simulated pileup interactions by $5\%$ to account for the
uncertainty in the luminosity and the uncertainty in the total inelastic cross section~\cite{CMS-PAPERS-FWD-11-001}, as determined by a comparison of the number of reconstructed vertices in $\Z\to\mu\mu$ events in data and simulation.
Uncertainties in the differences between data and simulation efficiencies of
the trigger, muon isolation, and muon identification criteria are
generally less than 1\%. An additional uncertainty due to the finite number of
simulated events used to construct the response matrix is calculated by randomly
varying the content of the response matrix according to a Poisson uncertainty in each bin.

\begin{figure}[htb]
\centering
\includegraphics[width=\cmsFigWidth]{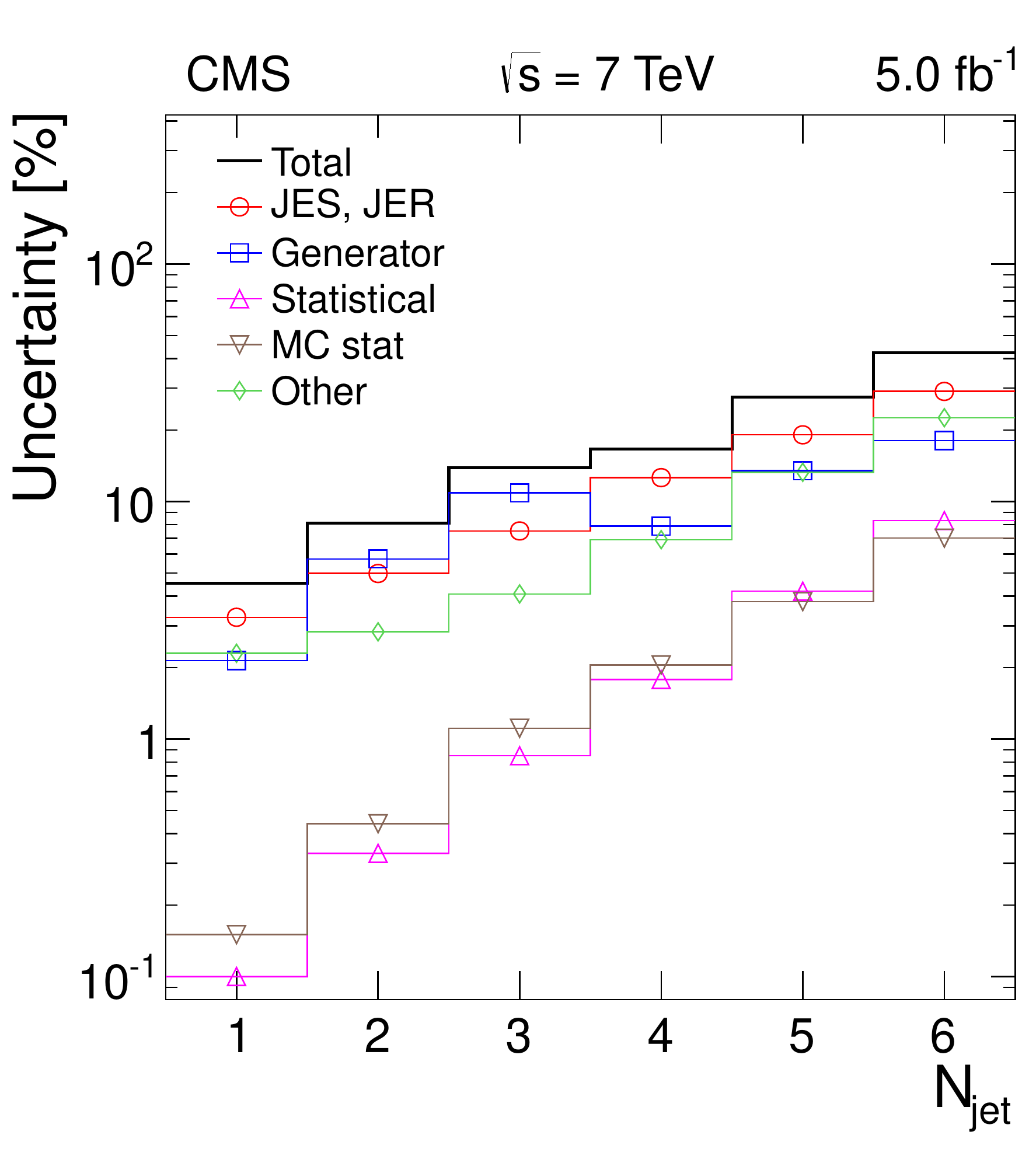}
\caption{The dominant systematic uncertainties in the measurement of the $\PW$+jets cross section as a function of the exclusive jet multiplicity. The systematic uncertainties displayed  include the jet energy scale and resolution (JES, JER), the choice of generator used in the unfolding procedure (Generator), the statistical uncertainty in the data minus the background, propagated through the unfolding procedure (Statistical), the uncertainty due to a finite number of simulated events used to construct the response matrix (MC stat.), and all other systematic uncertainties (Other) detailed in Section~\ref{systematics}, including pileup, integrated luminosity, background normalisation, b-tagging, muon momentum and resolution, trigger efficiency, muon identification. The uncertainties presented here correspond to the weighted average of the values shown in Table~\ref{tab:sysranges}.}
\label{fig:sysjetmult}

\end{figure}

The effect of the systematic variations on the measured cross section as a function of the exclusive jet multiplicity is illustrated in Fig.~\ref{fig:sysjetmult}. The uncertainties given in Fig.~\ref{fig:sysjetmult} are the total uncertainty for each jet multiplicity. The corresponding ranges of systematic uncertainty across bins of jet \pt are given in Table~\ref{tab:sysranges}.

\begin{table*}[htb]
\topcaption{Ranges of uncertainties for the measurement of $\rd\sigma/\rd\pt$ of the $n$th jet in events with $n$ or more jets. The uncertainties displayed include the statistical uncertainty in the data minus the background, propagated through the unfolding procedure (Statistical), the jet energy scale and resolution (JES, JER), the choice of generator used in the unfolding procedure (Generator), the uncertainty due to a finite number of simulated events used to construct the response matrix (MC stat.), and all other systematic uncertainties (Other) detailed in Section~\ref{systematics}, including pileup, integrated luminosity, background normalisation, b-tagging, muon momentum and resolution, trigger efficiency, muon identification.
}
\centering
\begin{tabular}{llcccccr}\hline
$n$ & \PT [GeV] & Statistical [\%]  & JES, JER [\%] & Generator [\%] & MC stat. [\%] & Other [\%] & Total [\%]  \\  \hline
1 & 30--850 &0.1--3.2 & 3.4--24  & 0.9--9.6  & 0.2--11   & 2.3--6.6  & 4.5--29 \\
2 & 30--550 &0.4--2.4  & 4.6--12  & 1.6--13 & 0.8--11  & 2.9--5.9  & 6.9--21 \\
3 & 30--450 &0.6--16  & 6.0--23  & 2.7--48 & 1.0--45 & 4.5--11 & 9.4--73 \\
4 & 30--210 &1.6--10 & 11--15 & 6.4--21 & 2.4--23  & 7.2--26 & 16--43 \\
\hline
\end{tabular}

\label{tab:sysranges}
\end{table*}

\section{Results}
\label{results}
\begin{figure*}[htb]
\centering
\includegraphics[width=\cmsFigWidthTwo]{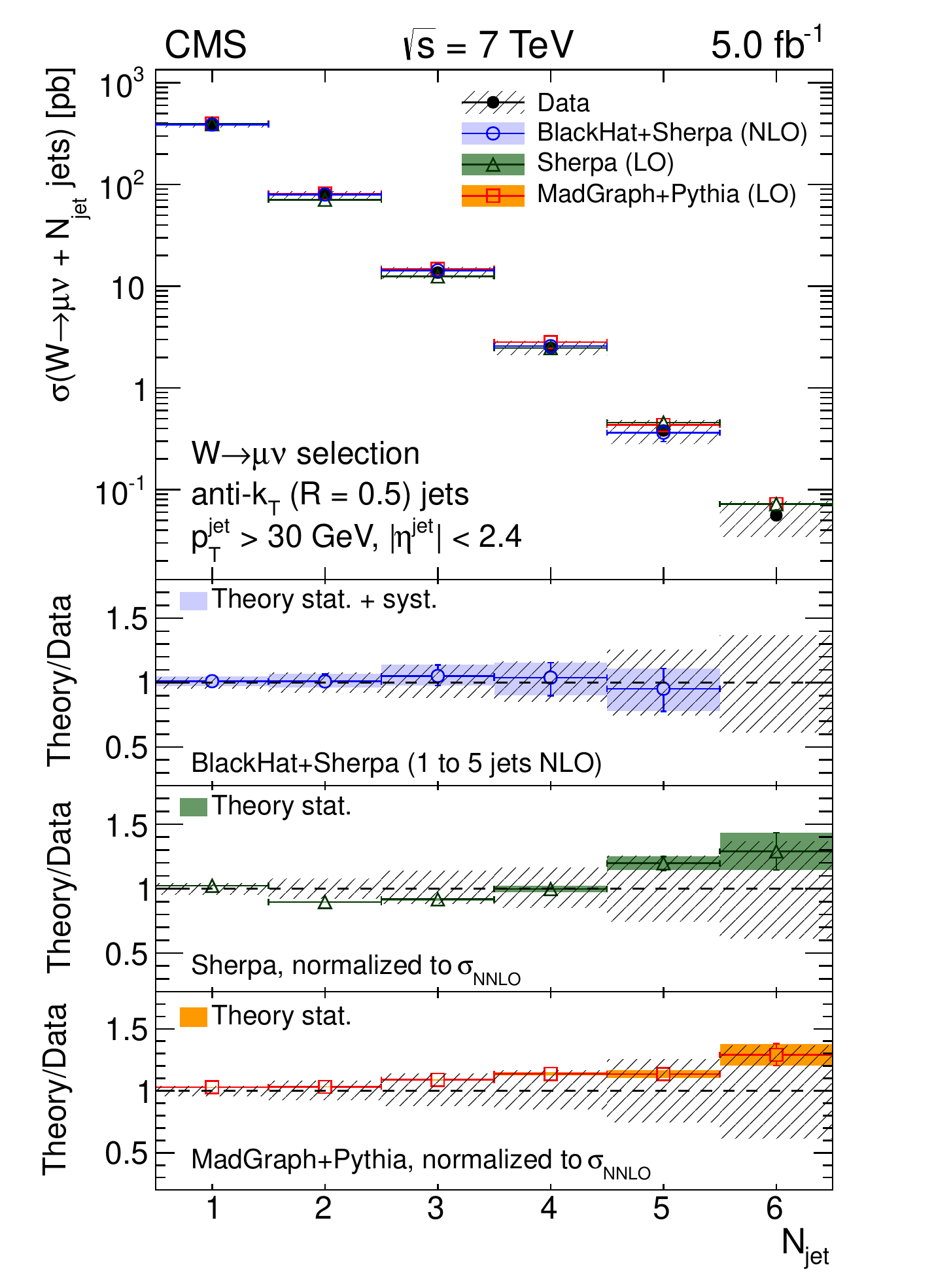}
\includegraphics[width=\cmsFigWidthTwo]{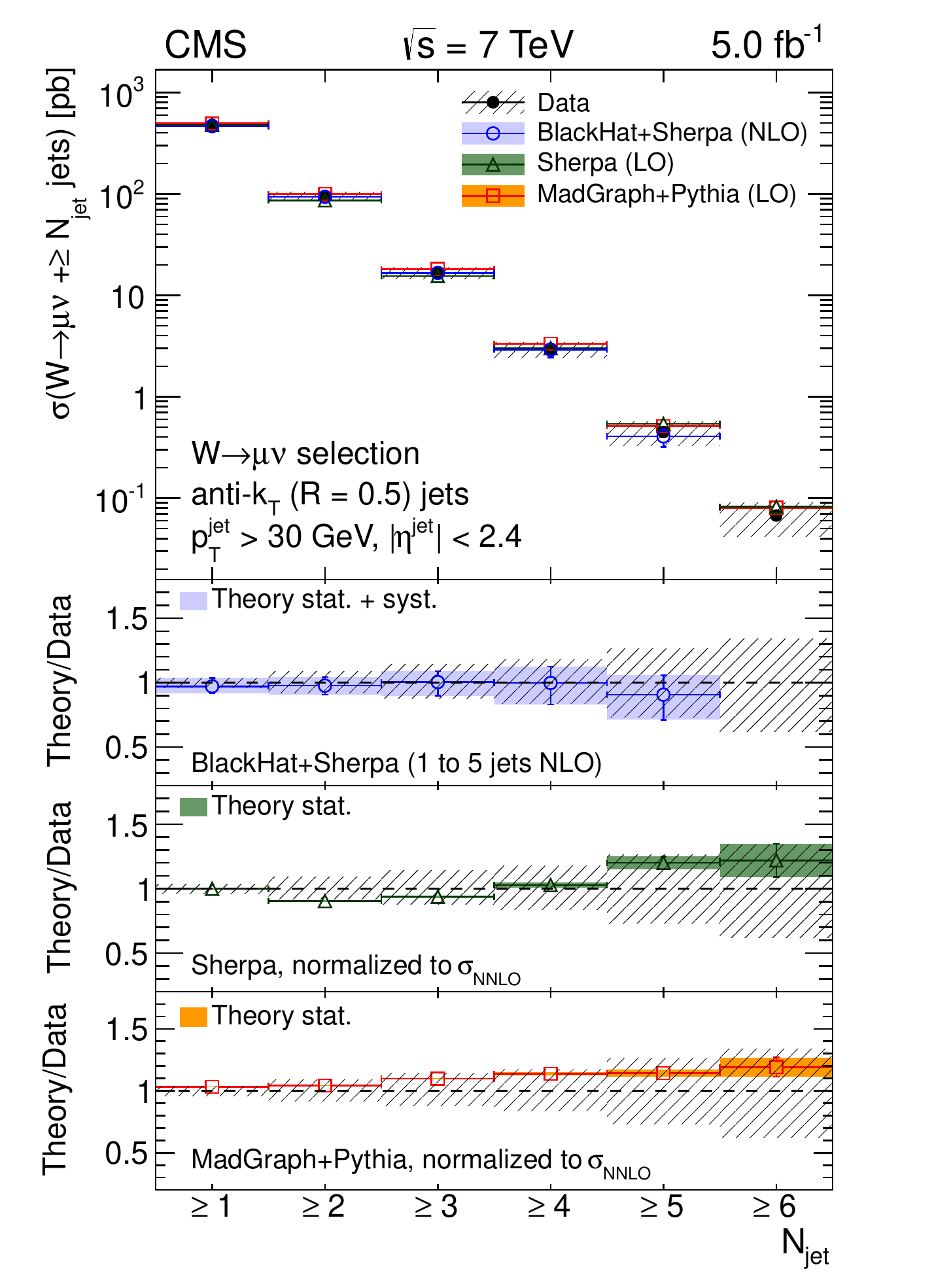}
\caption{The cross section measurement for the exclusive and inclusive jet multiplicities, compared to the predictions of \MADGRAPH 5.1.1 + \PYTHIA 6.426,  \SHERPA 1.4.0, and \BLACKHAT{}+\SHERPA (corrected for hadronisation and multiple-parton interactions). Black  circular markers with the grey hatched band represent the unfolded data measurement and its uncertainty. Overlaid are the predictions together with their statistical uncertainties (Theory stat.). The \BLACKHAT{}+\SHERPA uncertainty also contains theoretical systematic uncertainties (Theory syst.) described in Section~\ref{results}.  The lower plots show the ratio of each prediction to the unfolded data.}
\label{fig:pyplots/PFJet30CountFINAL_PlotXSec}

\end{figure*}

\begin{figure*}[htbp]
\centering
\includegraphics[width=\cmsFigWidthTall]{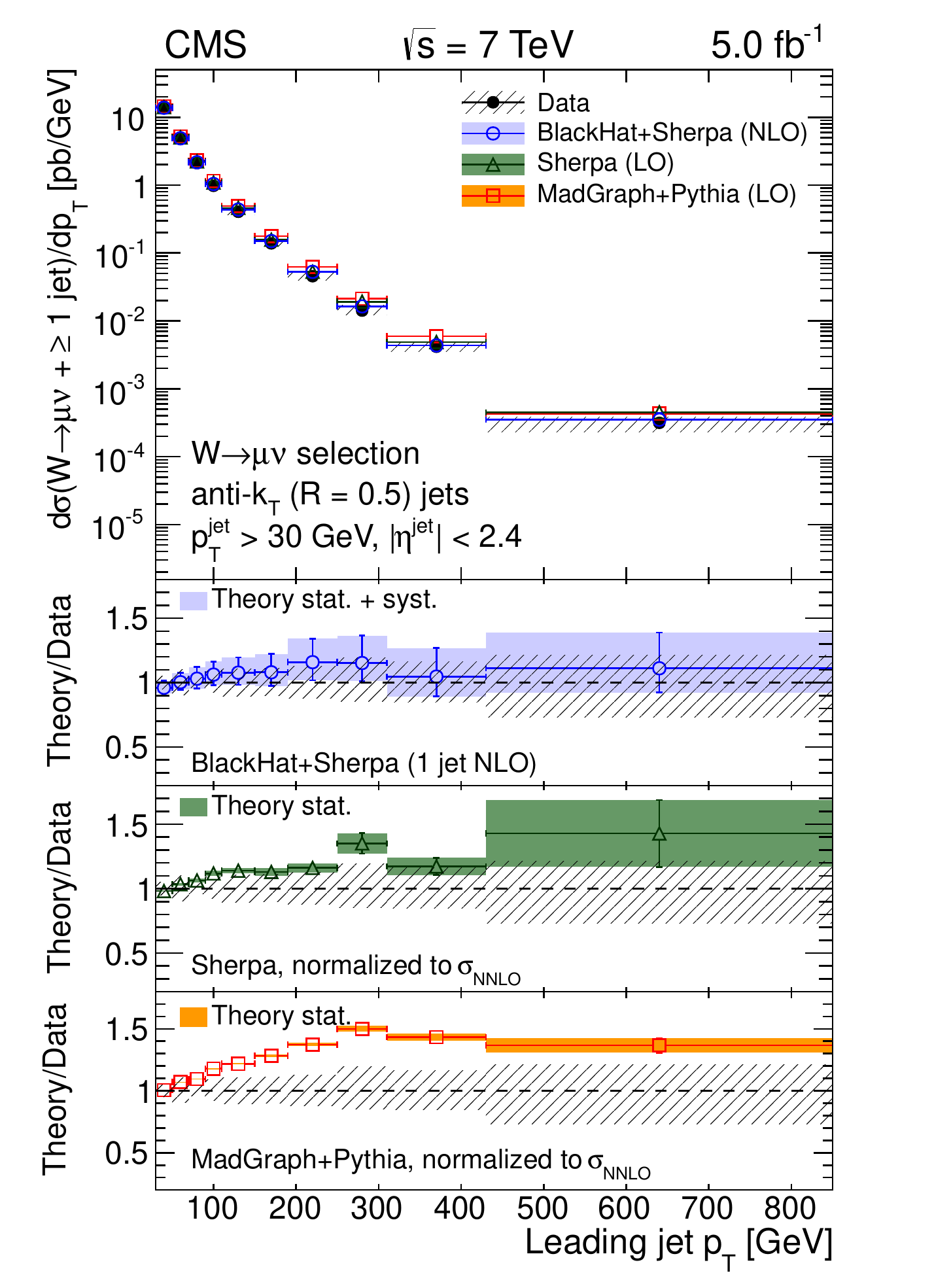}
\includegraphics[width=\cmsFigWidthTall]{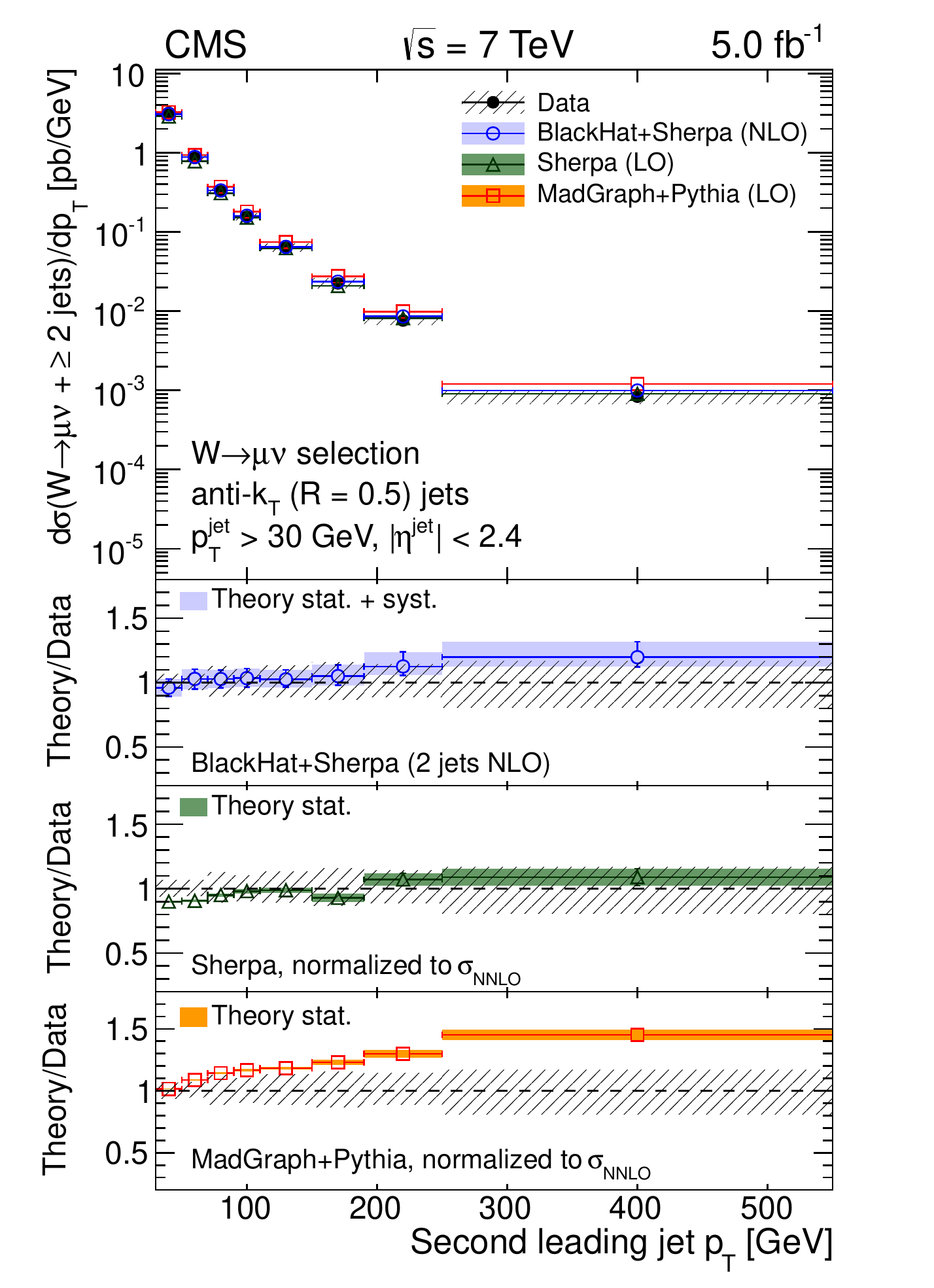}
\includegraphics[width=\cmsFigWidthTall]{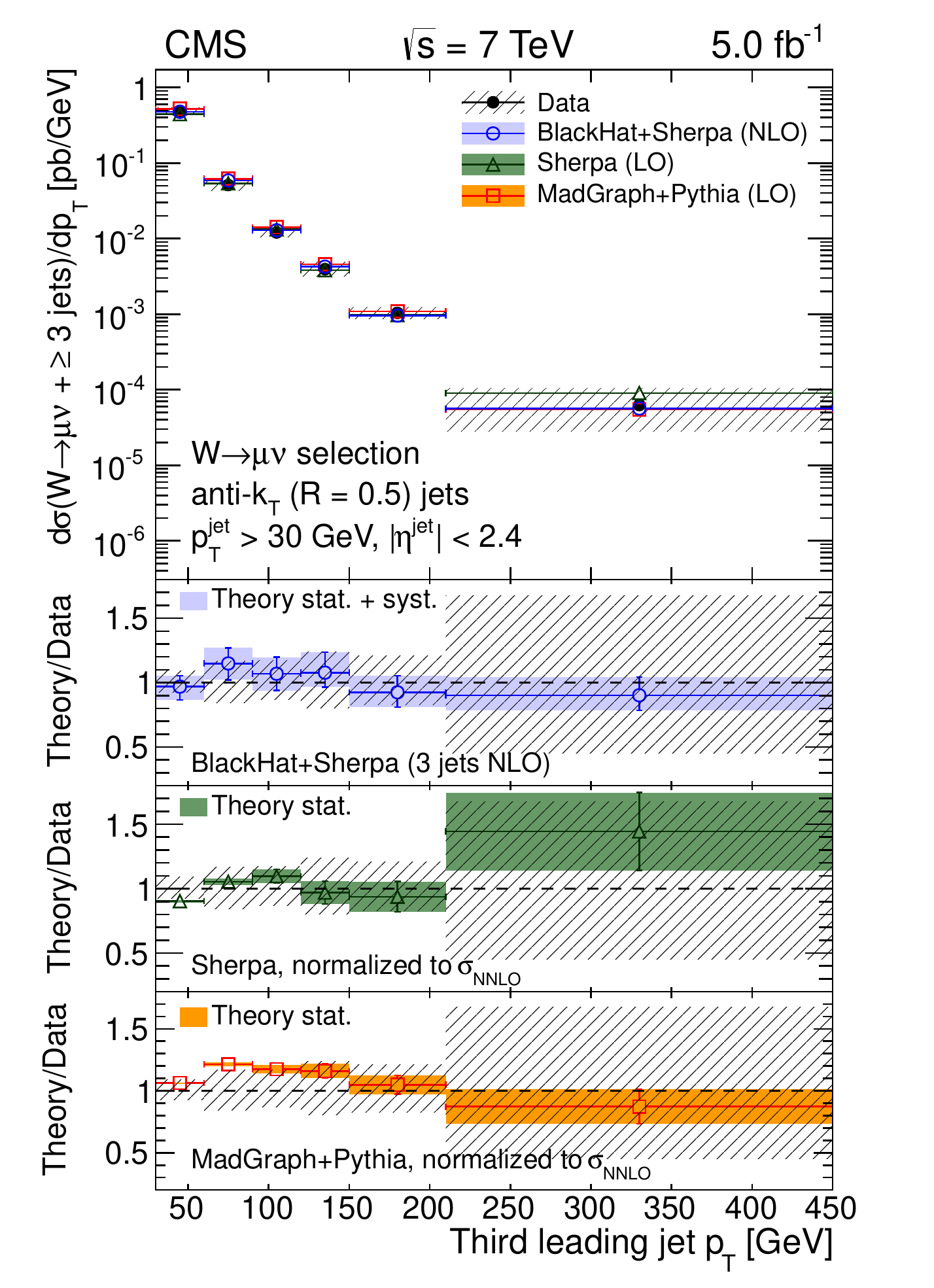}
\includegraphics[width=\cmsFigWidthTall]{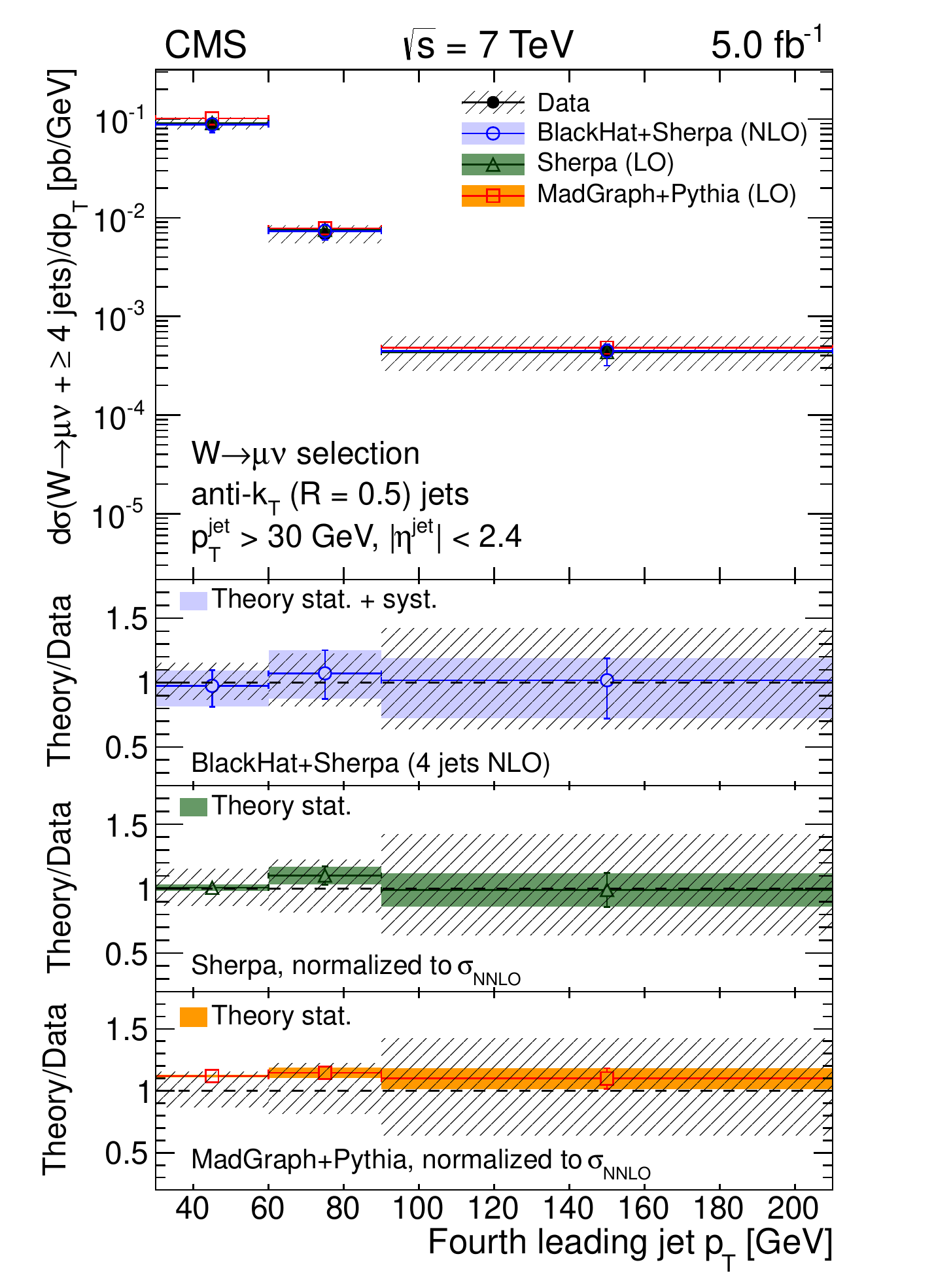}
\caption{The differential cross section measurement for the leading four jets' transverse momenta, compared to the predictions of \MADGRAPH 5.1.1 + \PYTHIA 6.426,  \SHERPA 1.4.0, and \BLACKHAT{}+\SHERPA (corrected for hadronisation and multiple-parton interactions).  Black  circular markers with the grey hatched band  represent the unfolded data measurement and its uncertainty. Overlaid are the predictions together with their statistical uncertainties (Theory stat.). The \BLACKHAT{}+\SHERPA uncertainty also contains theoretical systematic uncertainties (Theory syst.) described in Section~\ref{results}.  The lower plots show the ratio of each prediction to the unfolded data.}
\label{fig:pyplots/Pt_pfjetFINAL_PlotXSec}

\end{figure*}

\begin{figure*}[htbp]
\centering
\includegraphics[width=\cmsFigWidthTall]{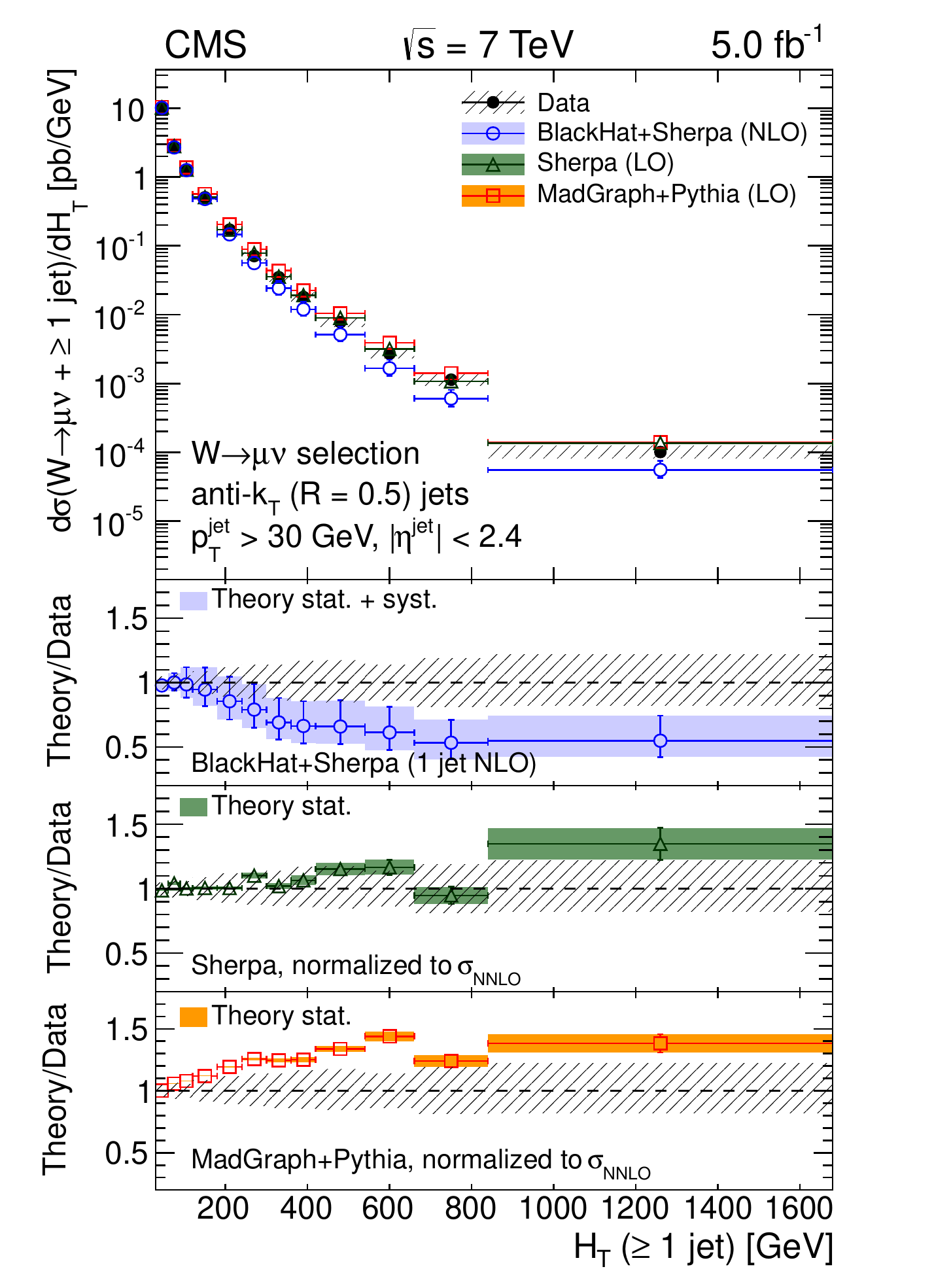}
\includegraphics[width=\cmsFigWidthTall]{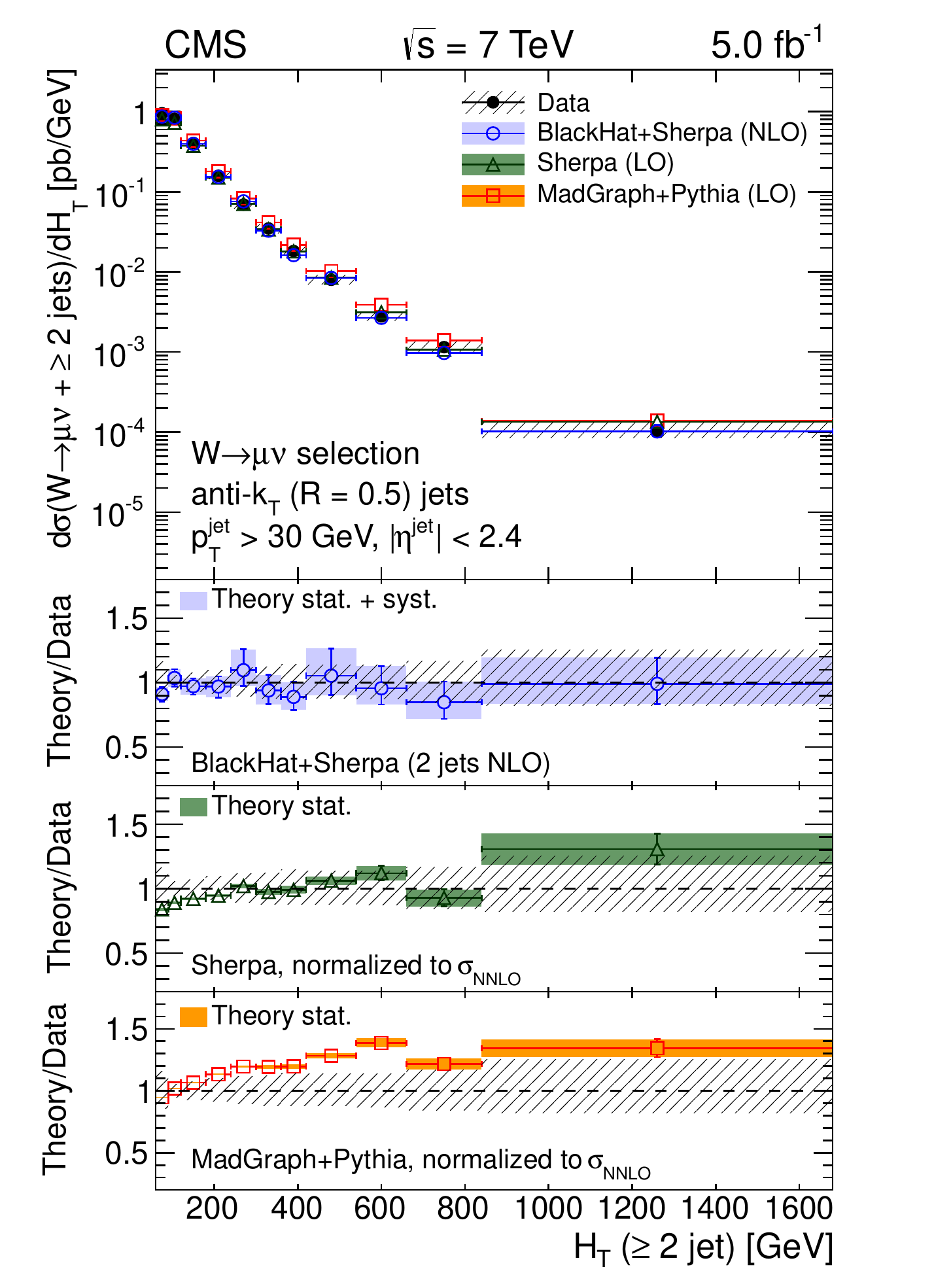}
\includegraphics[width=\cmsFigWidthTall]{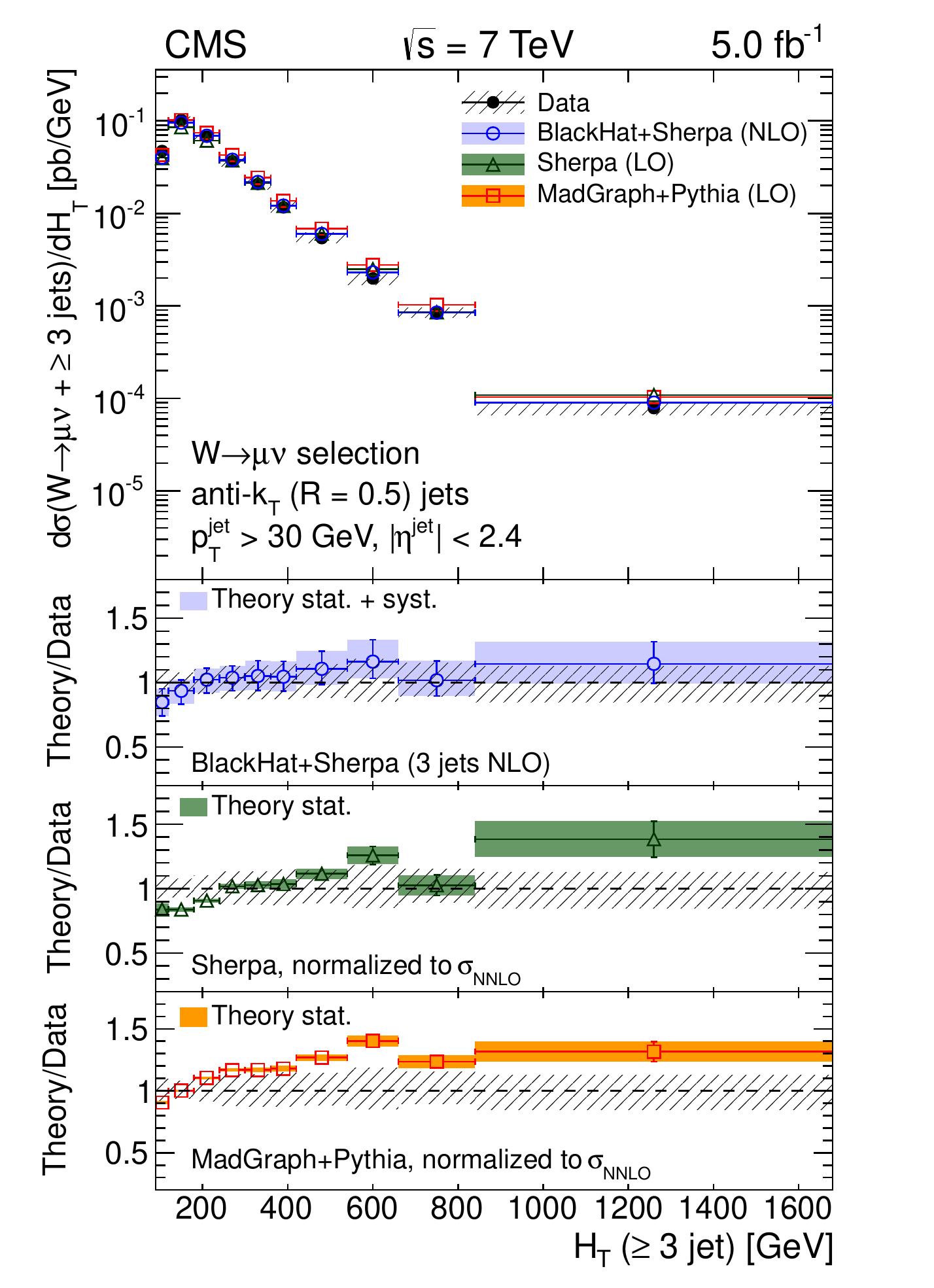}
\includegraphics[width=\cmsFigWidthTall]{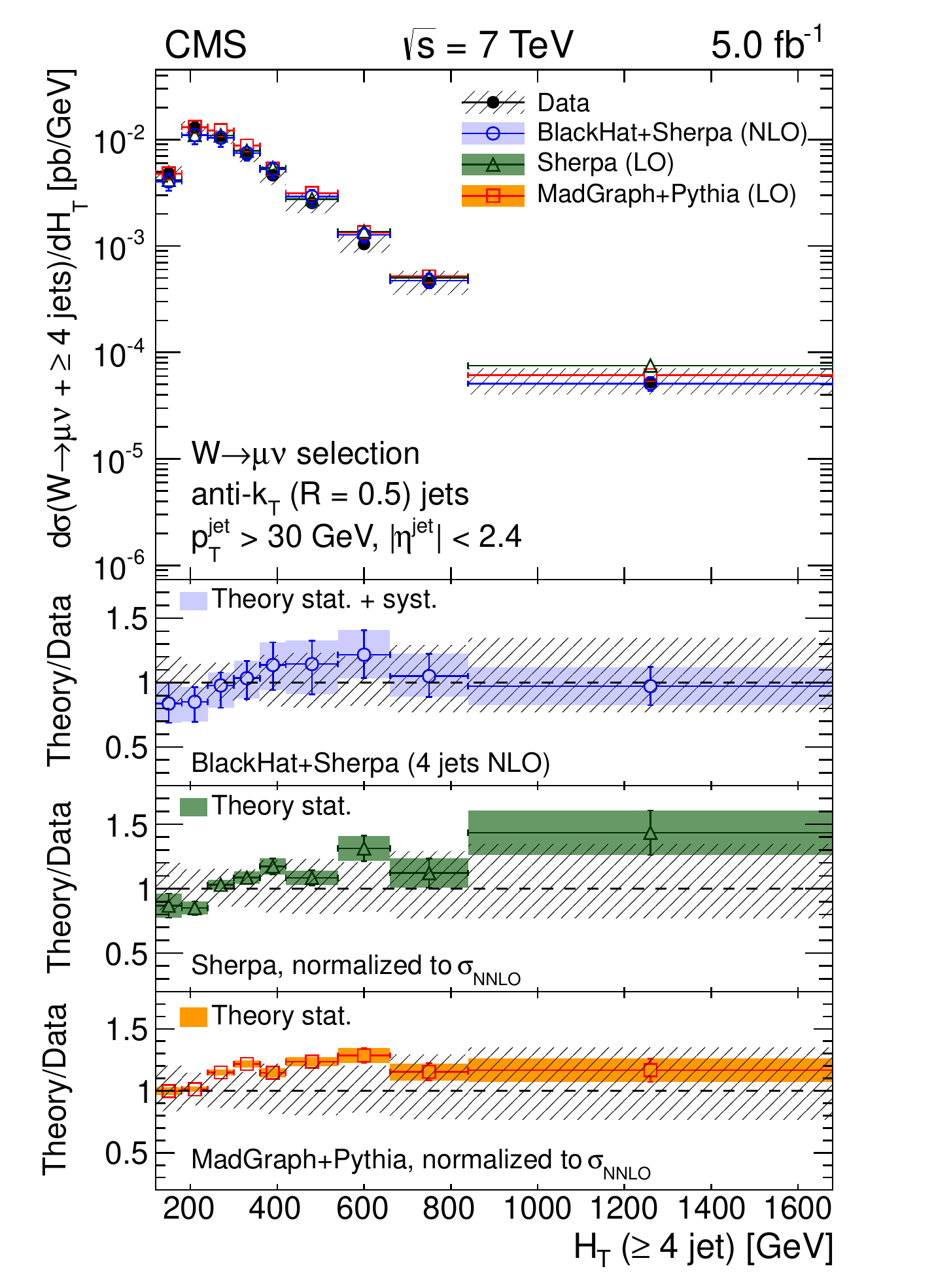}
\caption{The differential cross section measurement for \HT for inclusive jet multiplicities 1--4, compared to the predictions of \MADGRAPH 5.1.1 + \PYTHIA 6.426,  \SHERPA 1.4.0, and \BLACKHAT{}+\SHERPA (corrected for hadronisation and multiple-parton interactions). Black  circular markers with the grey hatched band represent the unfolded data measurement and its uncertainty. Overlaid are the predictions together with their statistical uncertainties (Theory stat.). The \BLACKHAT{}+\SHERPA uncertainty also contains theoretical systematic uncertainties (Theory syst.) described in Section~\ref{results}.  The lower plots show the ratio of each prediction to the unfolded data.}
\label{fig:pyplots/Ht_pfjetFINAL_PlotXSec}

\end{figure*}

\begin{figure*}[htbp]
\centering
\includegraphics[width=\cmsFigWidthTall]{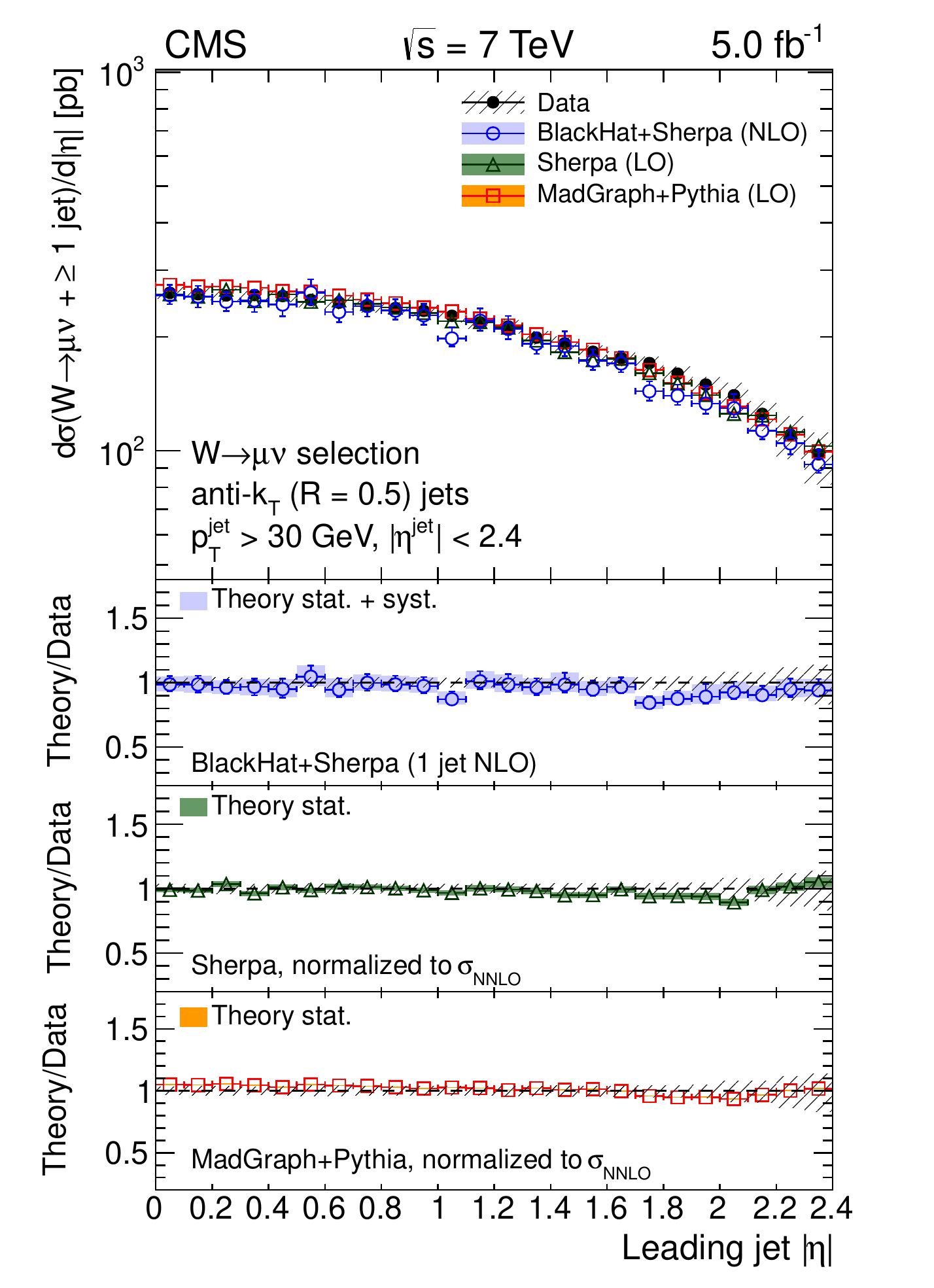}
\includegraphics[width=\cmsFigWidthTall]{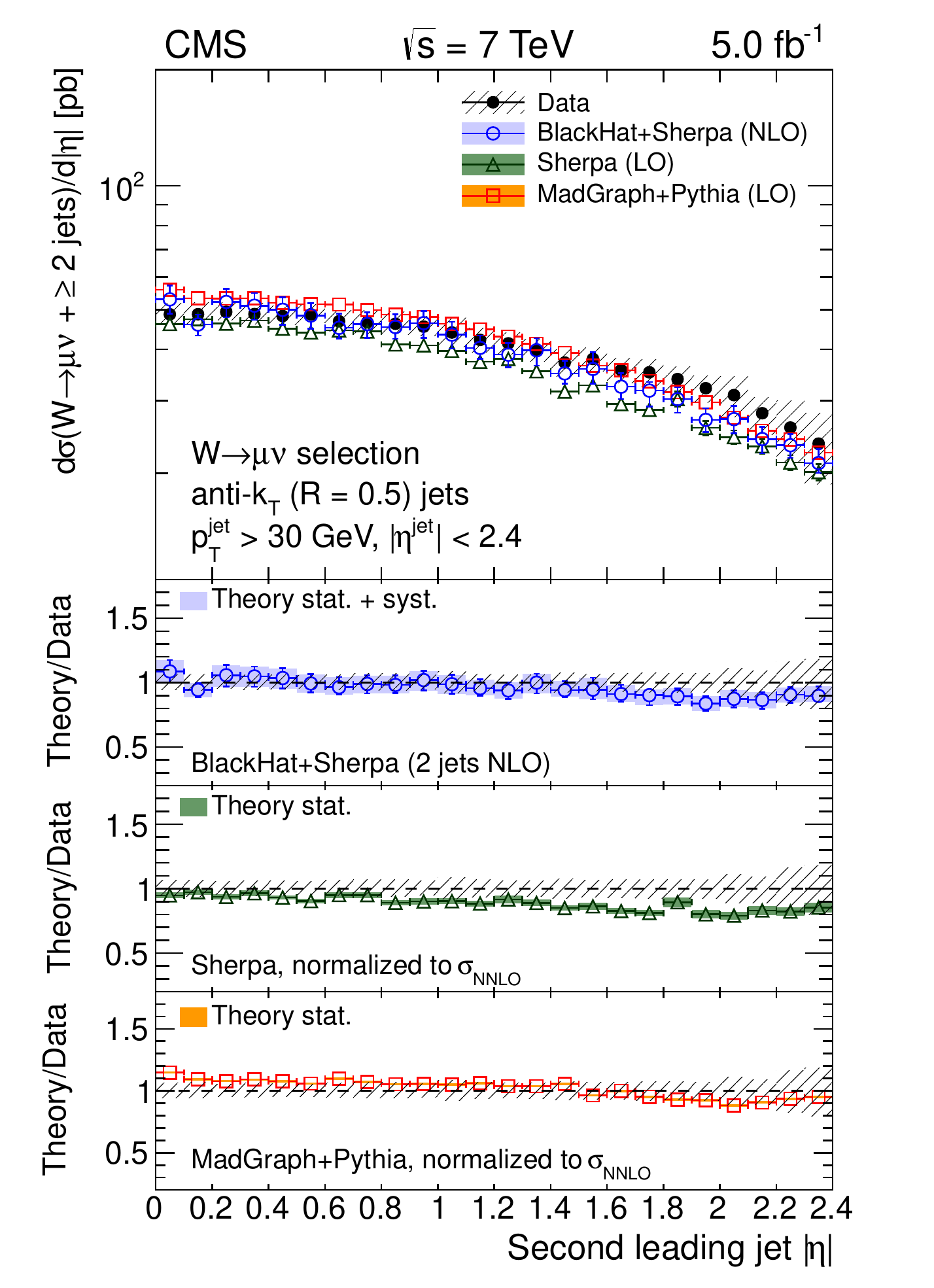}
\includegraphics[width=\cmsFigWidthTall]{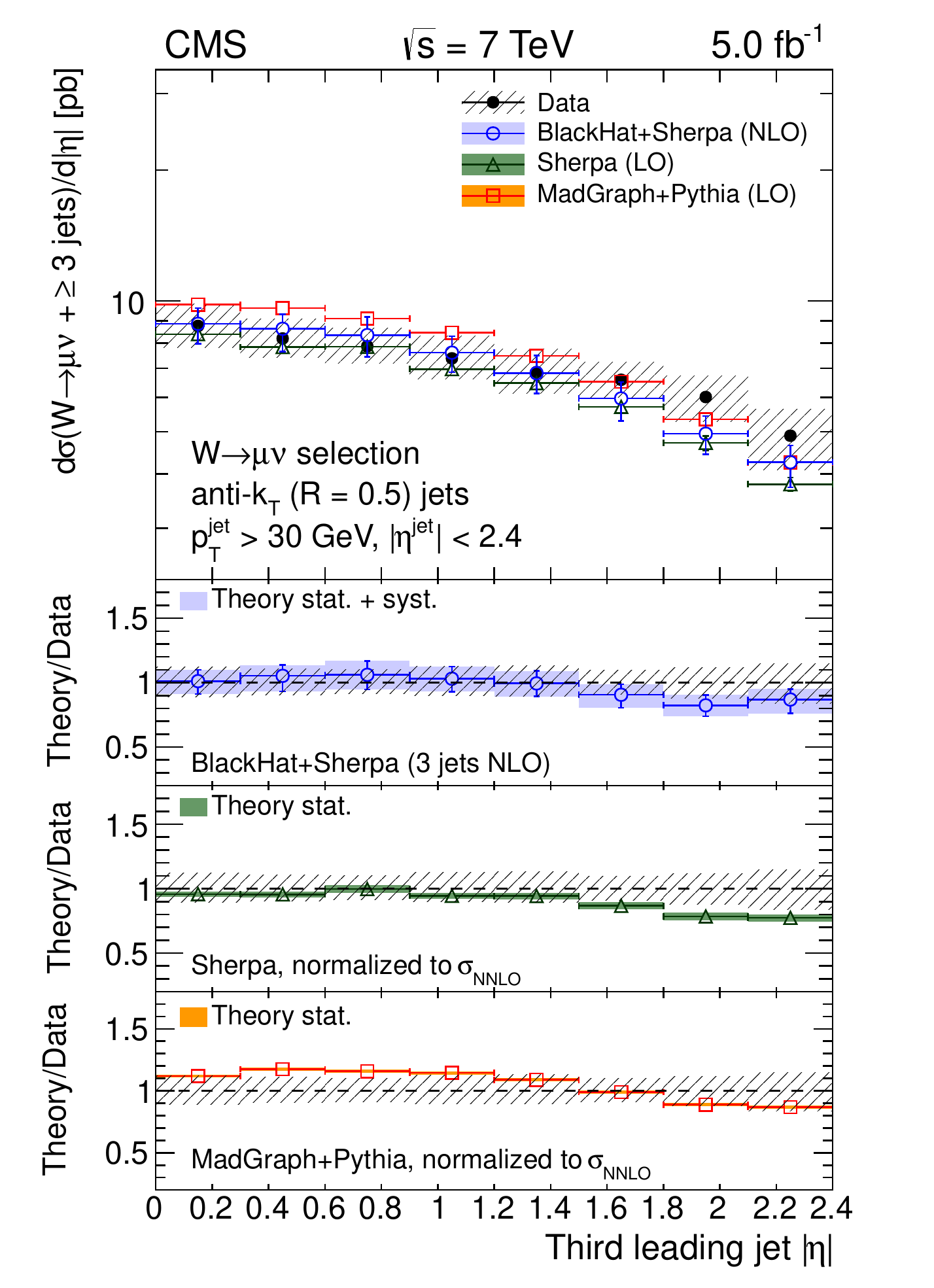}
\includegraphics[width=\cmsFigWidthTall]{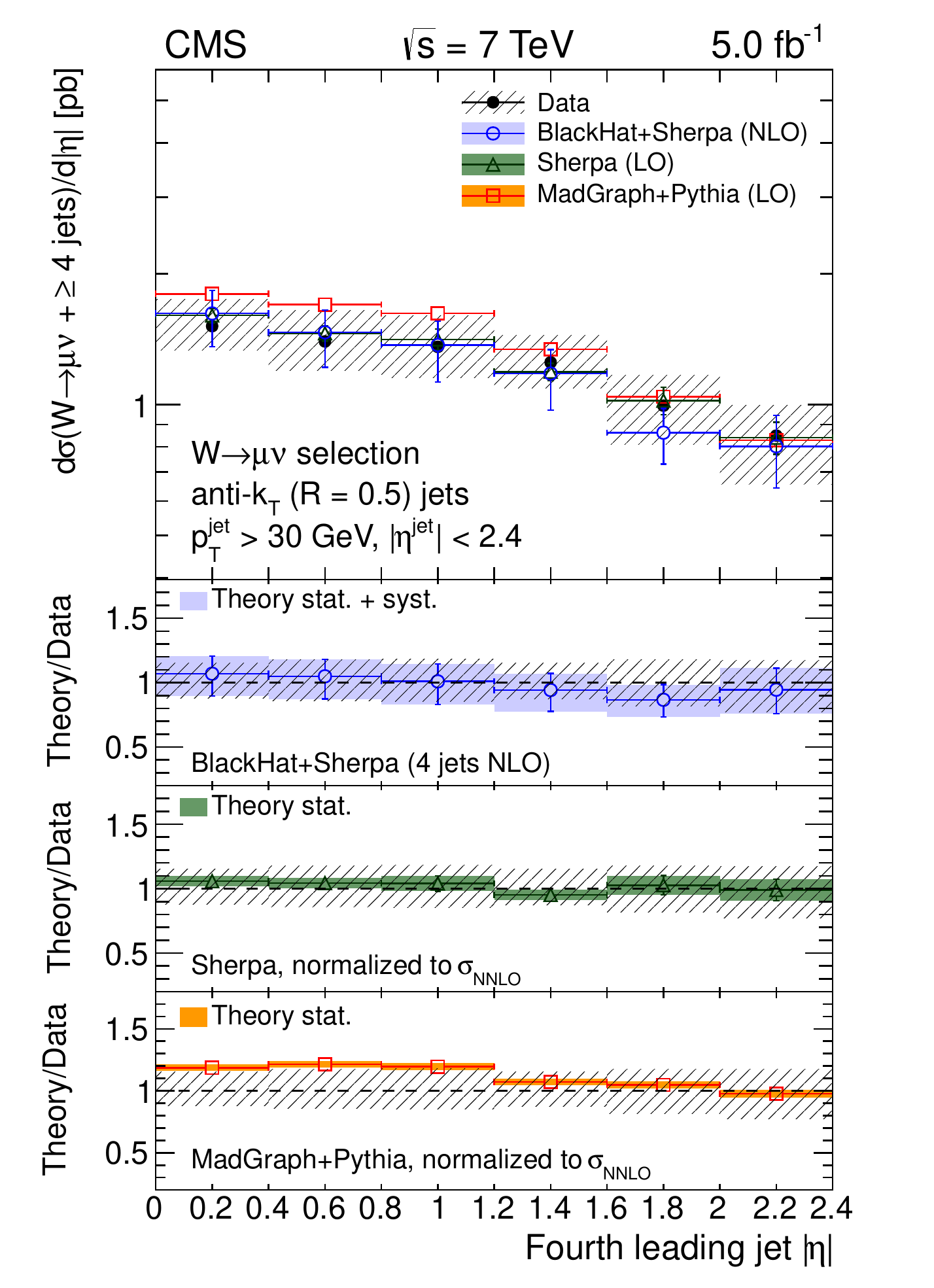}
\caption{The differential cross section measurement for the pseudorapidity of the four leading jets, compared to the predictions of \MADGRAPH 5.1.1 + \PYTHIA 6.426,  \SHERPA 1.4.0, and \BLACKHAT{}+\SHERPA (corrected for hadronisation and multiple-parton interactions).  Black  circular markers with the grey hatched band represent the unfolded data measurement and its uncertainty. Overlaid are the predictions together with their statistical uncertainties (Theory stat.). The \BLACKHAT{}+\SHERPA uncertainty also contains theoretical systematic uncertainties (Theory syst.) described in Section~\ref{results}.  The lower plots show the ratio of each prediction to the unfolded data.}
\label{fig:pyplots/Eta_pfjetFINAL_PlotXSec}

\end{figure*}

\begin{figure*}[htbp]
\centering
\includegraphics[width=\cmsFigWidthTall]{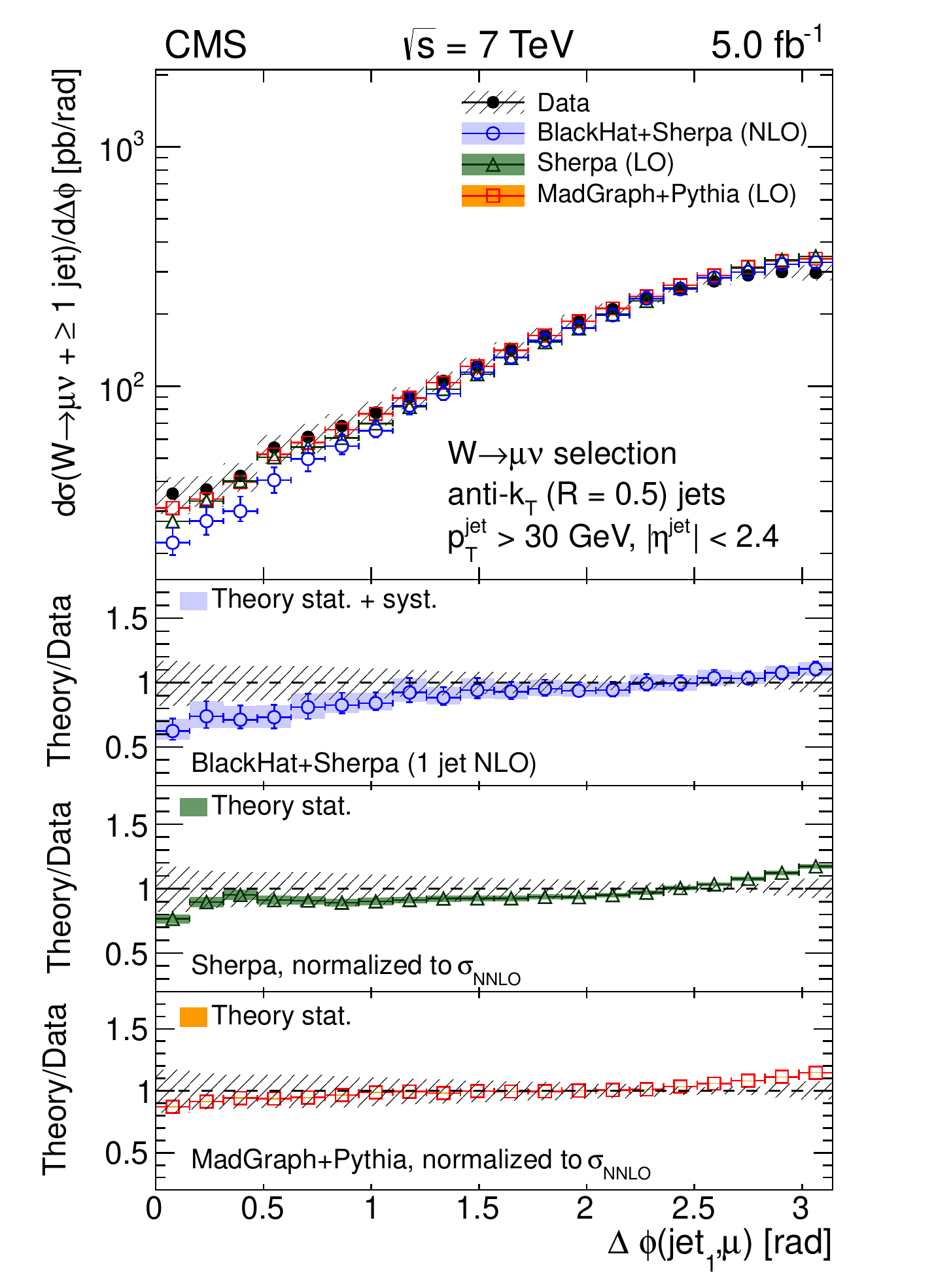}
\includegraphics[width=\cmsFigWidthTall]{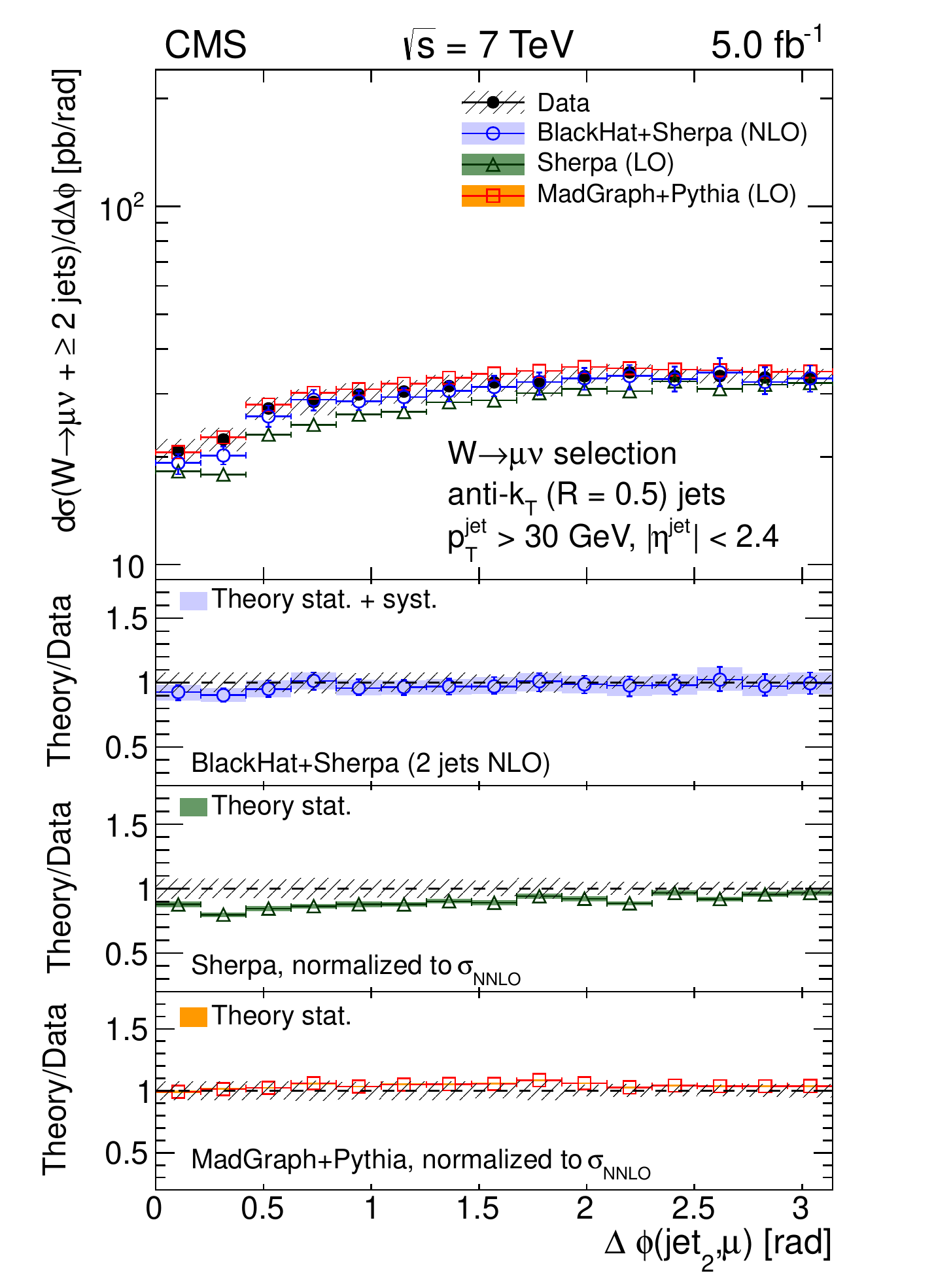}
\includegraphics[width=\cmsFigWidthTall]{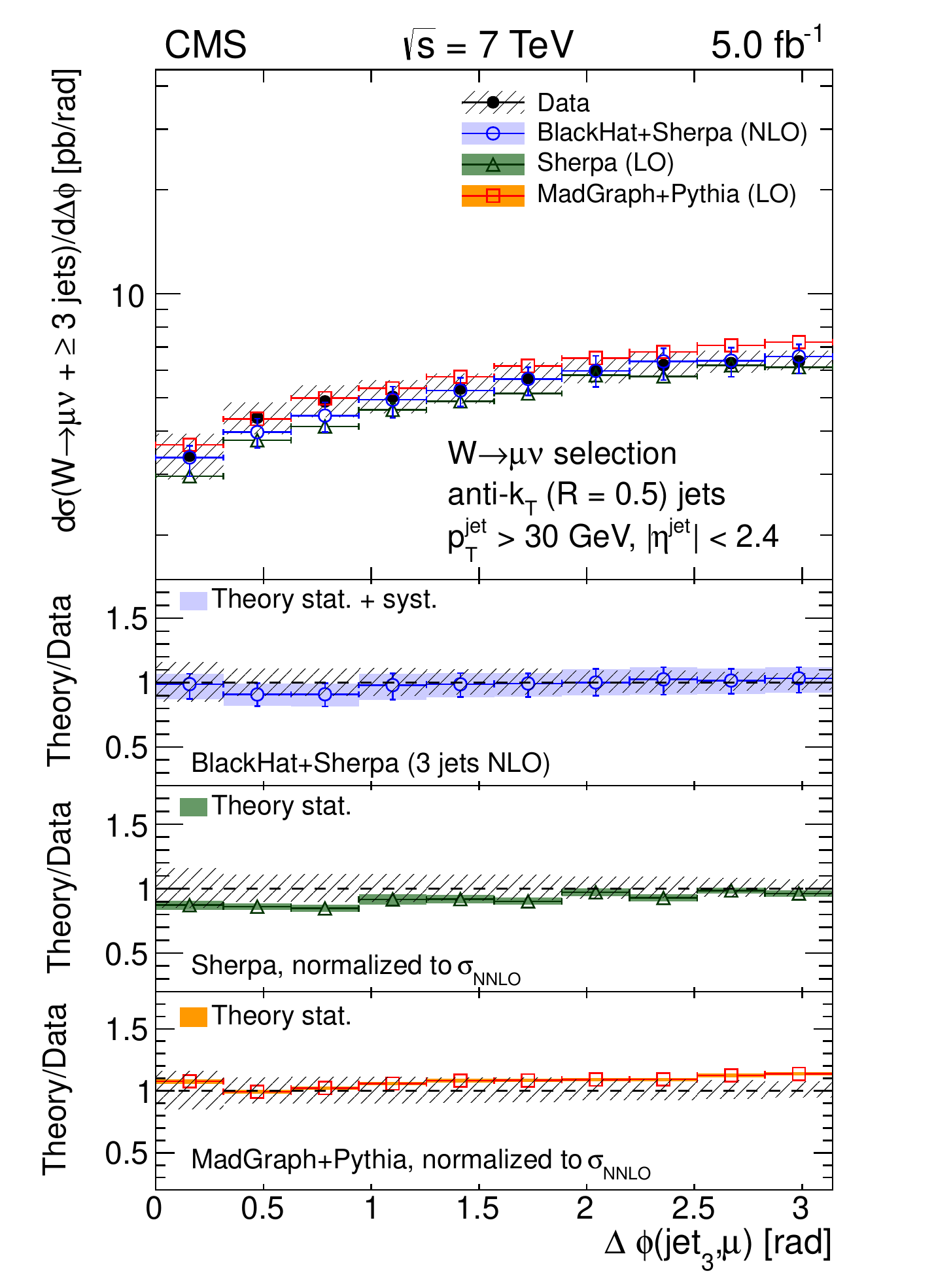}
\includegraphics[width=\cmsFigWidthTall]{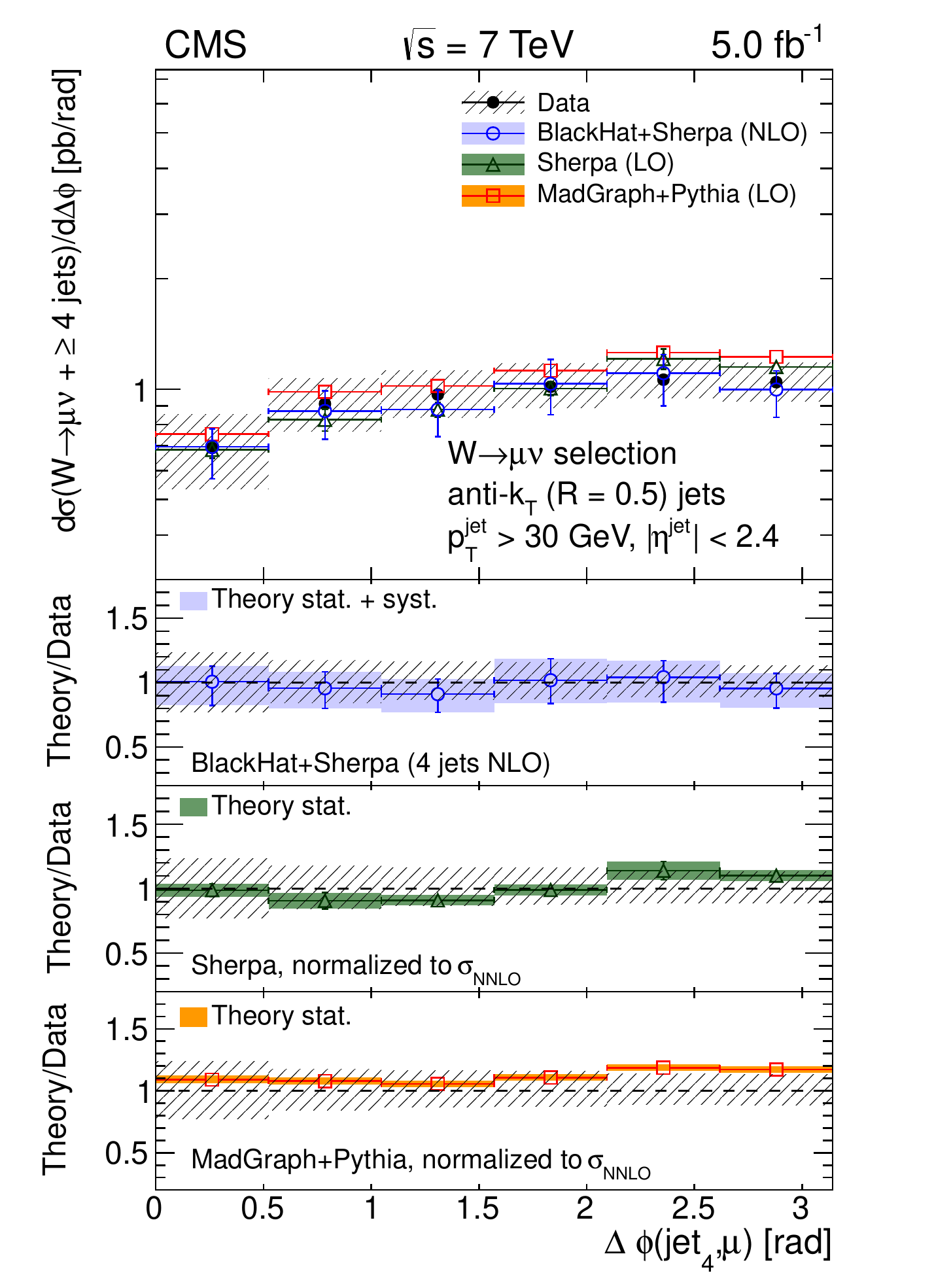}
\caption{The differential cross section measurement in $\Delta \phi (\text{jet}_n, \mu)$, for $n = 1$ - $4$, compared to the predictions of  \MADGRAPH 5.1.1 + \PYTHIA 6.426,  \SHERPA 1.4.0, and \BLACKHAT{}+\SHERPA (corrected for hadronisation and multiple-parton interactions).  Black  circular markers with the grey hatched band represent the unfolded data measurement and its uncertainty. Overlaid are the predictions together with their statistical uncertainties (Theory stat.). The \BLACKHAT{}+\SHERPA uncertainty also contains theoretical systematic uncertainties (Theory syst.) described in Section~\ref{results}.  The lower plots show the ratio of each prediction to the unfolded data.}
\label{fig:pyplots/DeltaPhi_pfjetmuon1FINAL_PlotXSec}

\end{figure*}

The cross sections for exclusive and inclusive jet multiplicities are given in Fig.~\ref{fig:pyplots/PFJet30CountFINAL_PlotXSec}. In Figs.~\ref{fig:pyplots/Pt_pfjetFINAL_PlotXSec}--~\ref{fig:pyplots/DeltaPhi_pfjetmuon1FINAL_PlotXSec} the differential cross sections are presented.
The measured $\PW$+jets cross sections are compared to the predictions
from several generators.
We consider $\PW$+jets signal processes generated with \MADGRAPH 5.1.1
using the CTEQ6L1 PDF set, with \SHERPA 1.4.0
 using the CT10~\cite{bib:CT10,bib:CT10N} PDF set, and with \BLACKHAT{}+\SHERPA~\cite{BLACKHAT} using the
CT10 PDF set.
Predictions from \MADGRAPH+\PYTHIA and \SHERPA are normalised to the  NNLO inclusive
cross sections calculated with \FEWZ~\cite{FEWZ}.
The \SHERPA sample is a separate sample from that used for the evaluation of uncertainties in Section~\ref{systematics}. The \MADGRAPH and \SHERPA predictions provide leading-order
(LO) matrix element (ME) calculations at each jet multiplicity, which are
then combined into inclusive samples by matching the ME
partons to particle jets. Parton showering (PS) and hadronisation of the \MADGRAPH
sample is performed with \PYTHIA 6.426
using the Z2 tune.
The \MADGRAPH+\PYTHIA calculation includes the production of up to four partons.
The jet matching is performed following the \kt-MLM prescription ~\cite{Matching}, where partons are clustered using the \kt algorithm with a distance parameter of $D=1$. The \kt clustering thresholds are chosen to be 10 GeV and 20 GeV at the matrix-element and parton-shower level, respectively.
The factorisation scale for each event is chosen to be the transverse mass computed after \kt-clustering of the event down to a 2$\to$2 topology. The renormalisation scale for the event is the \kt computed at each vertex splitting.
The predictions from \SHERPA include the production of up to four partons. The matching between jets and partons is performed with the CKKW matching scheme~\cite{Matching}, and the default factorisation and renormalisation scales are used.

The predictions from
\MADGRAPH{}+\PYTHIA and \SHERPA are shown with statistical uncertainties only.
These \MADGRAPH{}+\PYTHIA  and \SHERPA samples are processed
through the {\sc Rivet}
toolkit~\cite{rivet} in order to create particle level
distributions, which can be compared with the unfolded data.
The \BLACKHAT{}+\SHERPA samples represent fixed-order predictions at the level
of ME partons of $\PW+n$ jets at NLO accuracy, for $n=$ 1, 2, 3, 4, and 5 jets.
Each measured distribution for a given inclusive jet multiplicity
is compared with the corresponding fixed-order prediction from \BLACKHAT{}+\SHERPA.
The choice of renormalisation and factorisation scales for \BLACKHAT{}+\SHERPA is
$\hat{H}_\mathrm{T}^{'}/2$, where $\hat{H}_\mathrm{T}^{'} \equiv \sum_{m}^{} \pt^m + E_{\mathrm{T}}^\PW$, $m$ represents the final state partons, and $E_{\mathrm{T}}^\PW$ is the transverse energy of the \PW{}~boson.
Before comparing to data, a nonperturbative correction is applied
to the \BLACKHAT{}+\SHERPA distributions to account for the effects of
multiple-parton interactions and hadronisation.  The nonperturbative
correction is determined using \MADGRAPH 5.1.1 interfaced to
\PYTHIA 6.426 and turning on and off the hadronisation and multiple-parton
interactions. The magnitude of the nonperturbative correction is typically 1--5\%,
and is calculated for each bin of each measured distribution. The model dependence of the
nonperturbative correction is negligible~\cite{Bern:2011ep}.
The \BLACKHAT{}+\SHERPA prediction
also includes uncertainties due to the PDF and variations of the factorisation and renormalisation
scales.  The nominal prediction is given by the central value of the CT10
PDF set, and the PDF uncertainty considers the envelope of the error sets of CT10, MSTW2008nlo68cl~\cite{bib:MSTW2008NLO}, and NNPDF2.1~\cite{bib:NNPDF2.1} according to the PDF4LHC prescription~\cite{Alekhin:2011sk,pdf4lhc2}.  The factorisation and renormalisation scale uncertainty is determined by varying the scales simultaneously by a factor 0.5 or 2.0.

The unfolded exclusive and inclusive jet multiplicity distributions, shown in Fig.~\ref{fig:pyplots/PFJet30CountFINAL_PlotXSec}, are found to be in agreement, within uncertainties, with the predictions of the generators and with the NLO calculation of \BLACKHAT{}+\SHERPA.
Table~\ref{tab:pyplotsMODDEDTCHEM__PFJet30CountFINALPlotXSec} details the measured cross sections as a function of the inclusive and exclusive jet multiplicity.

The jet \pt unfolded distributions for inclusive jet multiplicities from 1 to 4 are shown in Fig.~\ref{fig:pyplots/Pt_pfjetFINAL_PlotXSec}. The predictions of \BLACKHAT{}+\SHERPA are in agreement with the measured distributions within the systematic uncertainties, while \MADGRAPH{}+\PYTHIA is observed to overestimate the yields up to 50\% (45\%) for the first (second) leading jet \pt distributions at high-\pt values. The predictions from \SHERPA are found to agree well for the second-, third-, and fourth-leading jet \pt distributions, while an excess of slightly more than one standard deviation can be seen at high-\pt values for the leading jet \pt distribution. Similar observations hold for \MADGRAPH{}+\PYTHIA and \SHERPA predictions in the \HT distributions for inclusive jet multiplicities of 1--4, as shown in Fig.~\ref{fig:pyplots/Ht_pfjetFINAL_PlotXSec}. Since the \BLACKHAT{}+\SHERPA NLO prediction for $\HT({\geq}1\:\text{jet})$ is a fixed-order prediction with up to two real partons, contributions from higher jet multiplicities are missing, which results in an underestimation in the tail of the distribution~\cite{Maitre:2013wha}. Similar observations have been made with $\PW$+jets measurements at D0~\cite{PhysRevD.88.092001} and ATLAS~\cite{ATLAS36pbWJets}.
In general, \SHERPA models the \HT distributions better than other generators.

The distributions of the jet $\eta$ and of the difference in azimuthal angle between each jet and the muon are shown in Figs.~\ref{fig:pyplots/Eta_pfjetFINAL_PlotXSec}~and~\ref{fig:pyplots/DeltaPhi_pfjetmuon1FINAL_PlotXSec}, respectively. The measurements of the jet $\eta$ agree with predictions from all generators, with {\sc MadGraph+Pythia} and  \BLACKHAT{}+\SHERPA performing best. The measurements of the $\Delta \phi$ between the leading jet and the muon are underestimated by as much as 38\% by \BLACKHAT{}+\SHERPA, with similar, but smaller, underestimations in predictions from \MADGRAPH{}+\PYTHIA and \SHERPA.

\begin{figure*}[htb]
\centering
\includegraphics[width=\cmsFigWidthTwo]{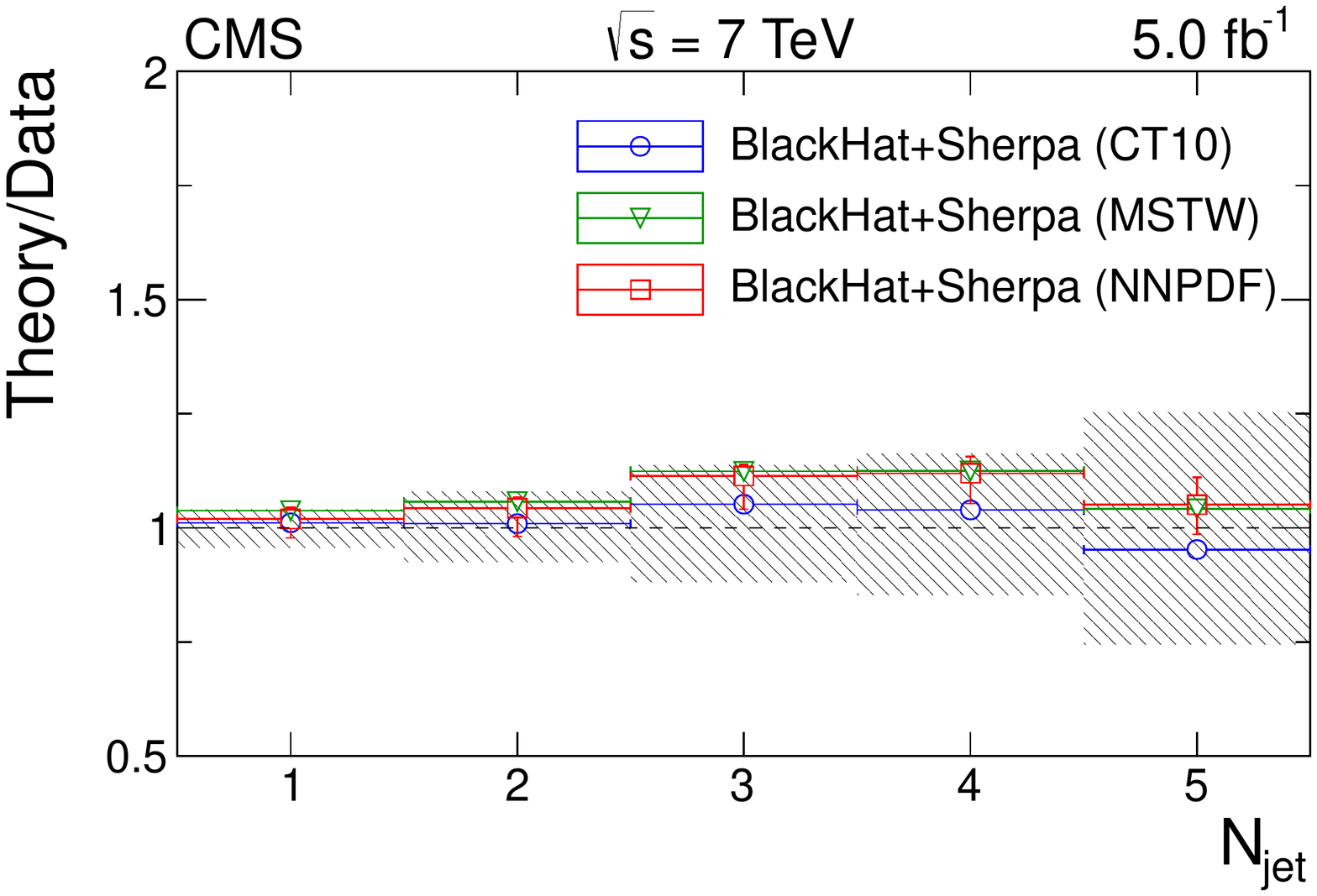}
\includegraphics[width=\cmsFigWidthTwo]{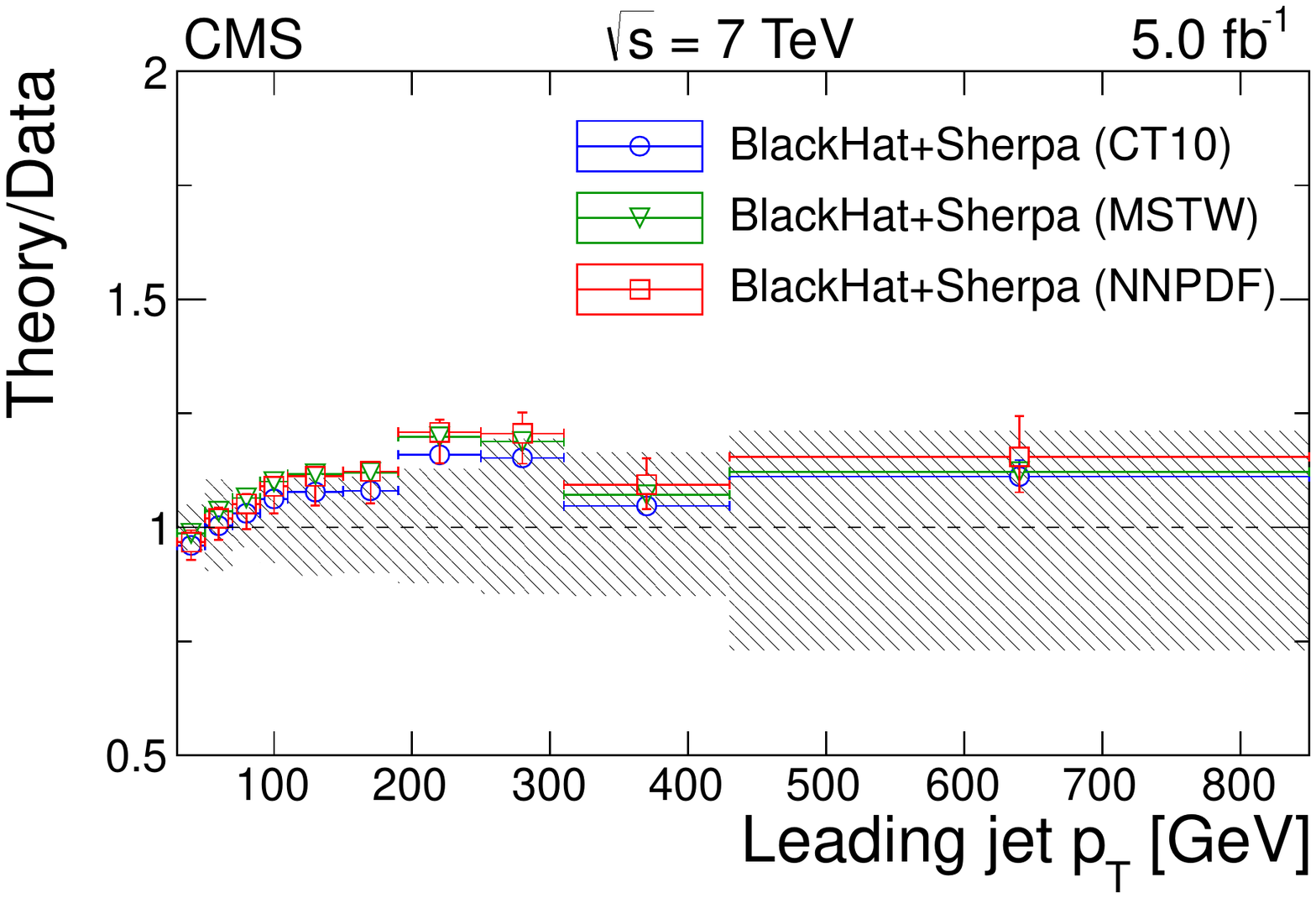}\\
\includegraphics[width=\cmsFigWidthTwo]{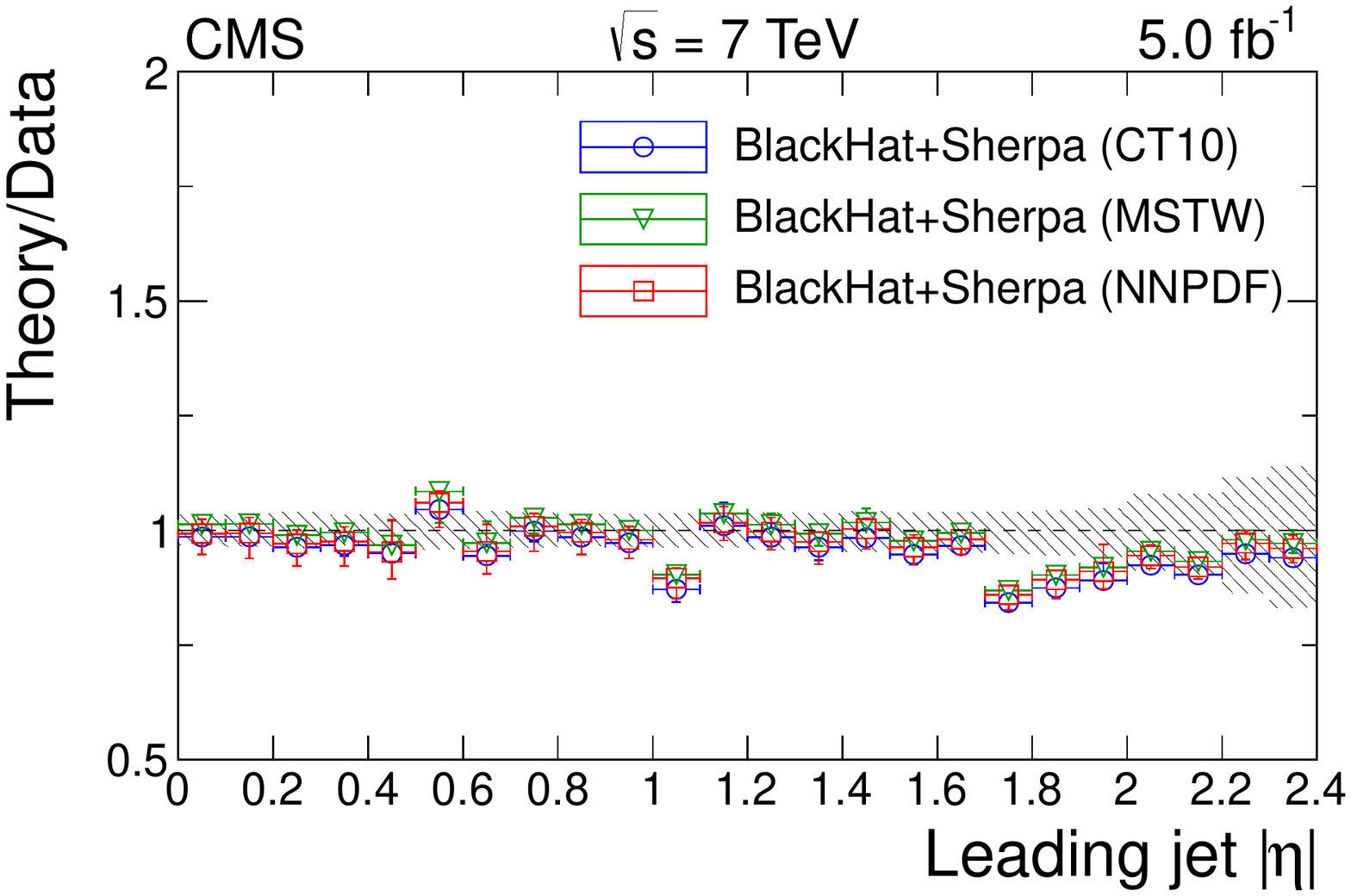}
\includegraphics[width=\cmsFigWidthTwo]{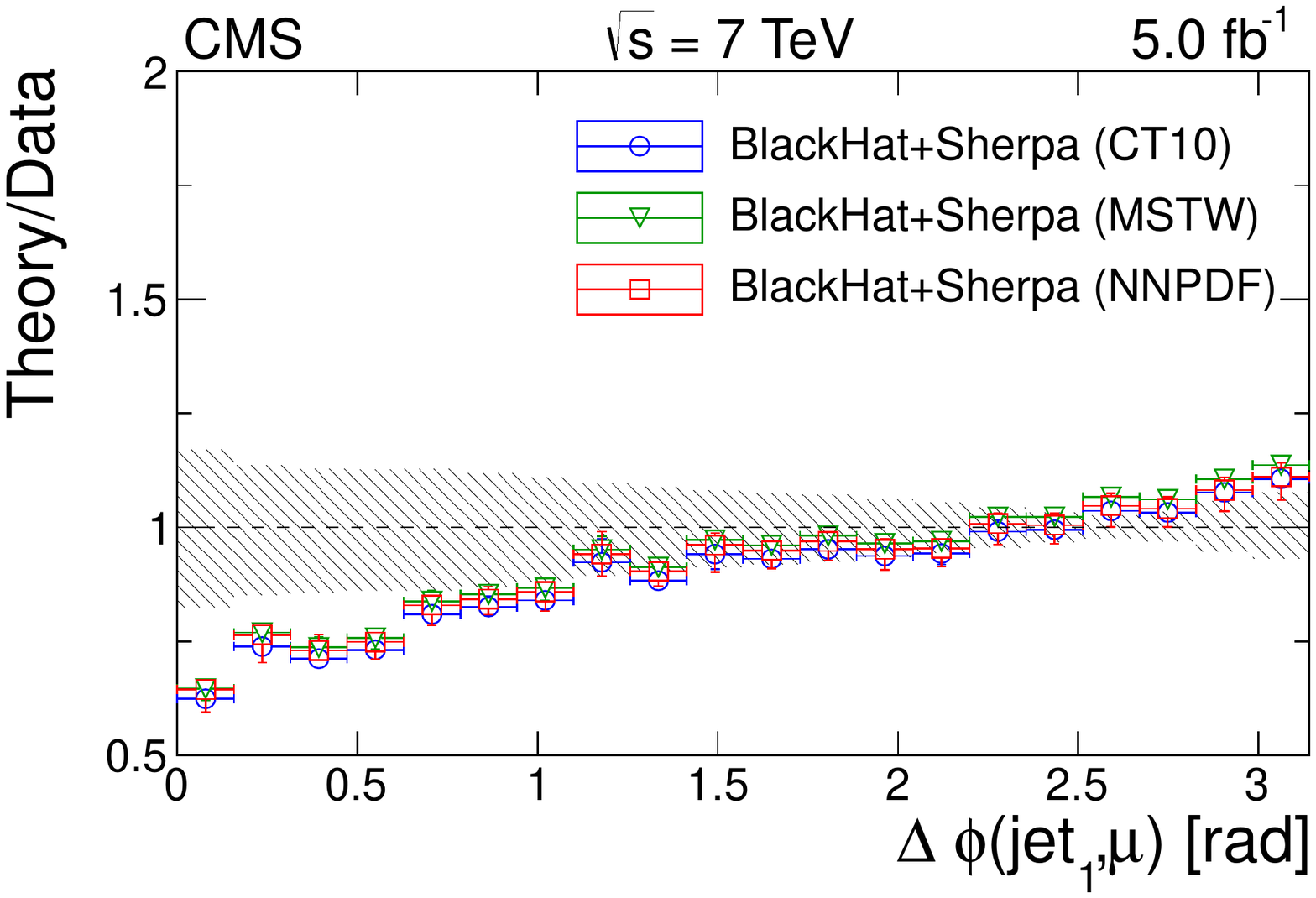}
\caption{The ratio of the predictions of \BLACKHAT{}+\SHERPA to the cross section measurements for four different quantities. The circular, triangular, and square markers indicate the predictions using the CT10, MSTW2008nlo68cl, and NNPDF PDF sets, respectively. The grey hatched band indicates the total uncertainty in the unfolded data measurement.}
\label{fig:PFJet30CountFINAL_PlotXSec_pdfcomparison}

\end{figure*}

Examples of the variation in the \BLACKHAT{}+\SHERPA prediction due to the choice of PDF are
given in Fig.~\ref{fig:PFJet30CountFINAL_PlotXSec_pdfcomparison}, in which the predictions with the MSTW2008nlo68cl, NNPDF2.1, and CT10 PDF sets are compared to the measurements from data.
The distributions determined with the different PDF sets are consistent with one another.

\begin{table}[htb]
\topcaption{Cross section measurements with statistical and systematic uncertainties for inclusive and exclusive jet multiplicities up to 6 jets. }
\centering
{\renewcommand{\arraystretch}{1.33}%
\begin{tabular}{cll}\hline
Jet multiplicity & Exclusive $\sigma$ [pb] & Inclusive $\sigma$ [pb]  \\  \hline \relax
1 & 384$^{+15}_{-17}$         & 480$^{+18}_{-20}$          \\
2 & 79.1$^{+6.2}_{-5.9}$      & 95.6$^{+8.5}_{-8.0}$       \\
3 & 13.6$^{+1.9}_{-1.6}$      & 16.6$^{+2.3}_{-2.0}$       \\
4 & 2.48$^{+0.40}_{-0.36}$    & 2.93$^{+0.52}_{-0.48}$     \\
5 & 0.382$^{+0.097}_{-0.097}$ & 0.45$^{+0.12}_{-0.12}$  \\
6 & 0.056$^{+0.020}_{-0.022}$ & 0.067$^{+0.023}_{-0.026}$  \\
\hline
\end{tabular}}

\label{tab:pyplotsMODDEDTCHEM__PFJet30CountFINALPlotXSec}
\end{table}

\section{Summary}
\label{conclusion}

Measurements of the cross sections and differential cross sections for a $\PW$ boson produced in association with jets in pp collisions at a centre-of-mass energy of 7\TeV have been presented.
The data were collected with the CMS detector
during the 2011 pp run of the LHC, and correspond to
an integrated luminosity of 5.0\fbinv. Cross sections have been
determined using the muon decay mode of the W boson and were presented
as functions of the jet multiplicity, the transverse momenta and
pseudorapidities of the four leading jets, the difference in
azimuthal angle between each jet and the muon, and the $\HT$ for jet multiplicities up to four. The results, corrected
for all detector effects by means of regularised unfolding, have been
compared with particle-level simulated predictions from pQCD.

Predictions from generators, \MADGRAPH{}+\PYTHIA and  \SHERPA, and NLO calculations from
\BLACKHAT{}+\SHERPA, describe the jet multiplicity within the uncertainties.
The cross section as a function of the \pt of the leading jet is overestimated by \MADGRAPH{}+\PYTHIA and \SHERPA, especially at high-\pt. Some overestimation from \MADGRAPH{}+\PYTHIA can also be observed in the second- and third-leading jet \pt distributions. The cross sections as a function of \pt predicted by \BLACKHAT{}+\SHERPA  agree with the measurements within uncertainties.
The predictions from \BLACKHAT{}+\SHERPA  underestimate the measurement of the cross section as a function of \HT for $N_\text{jet} \geq 1$, since the contribution from $\PW$+$\geq$3 jets is missing from an NLO prediction of $\PW$+$\geq$1 jet.
The cross sections as a function of \HT for $N_\text{jet} \geq 2$, 3, and 4 predicted by \BLACKHAT{}+\SHERPA agree with the measurements within the uncertainties.
The distributions of $\Delta \phi$ between the leading jet and the muon are underestimated by all predictions for $\Delta \phi$ values near zero,  with the largest disagreement visible in \BLACKHAT{}+\SHERPA. The distributions of $\Delta \phi$ between the second-, third-, and fourth-leading jets and the muon agree with all predictions within uncertainties. No significant disagreement was found in the distributions of $\eta$ of the four leading jets.

\section*{Acknowledgments}

We extend our thanks to Daniel Ma\^{i}tre and Zvi Bern for the production
of data and tools used to create the \BLACKHAT{}+\SHERPA predictions, and for the sharing
of expertise and advice regarding these predictions.

We congratulate our colleagues in the CERN accelerator departments for the excellent performance of the LHC and thank the technical and administrative staffs at CERN and at other CMS institutes for their contributions to the success of the CMS effort. In addition, we gratefully acknowledge the computing centres and personnel of the Worldwide LHC Computing Grid for delivering so effectively the computing infrastructure essential to our analyses. Finally, we acknowledge the enduring support for the construction and operation of the LHC and the CMS detector provided by the following funding agencies: BMWFW and FWF (Austria); FNRS and FWO (Belgium); CNPq, CAPES, FAPERJ, and FAPESP (Brazil); MES (Bulgaria); CERN; CAS, MoST, and NSFC (China); COLCIENCIAS (Colombia); MSES and CSF (Croatia); RPF (Cyprus); MoER, ERC IUT and ERDF (Estonia); Academy of Finland, MEC, and HIP (Finland); CEA and CNRS/IN2P3 (France); BMBF, DFG, and HGF (Germany); GSRT (Greece); OTKA and NIH (Hungary); DAE and DST (India); IPM (Iran); SFI (Ireland); INFN (Italy); NRF and WCU (Republic of Korea); LAS (Lithuania); MOE and UM (Malaysia); CINVESTAV, CONACYT, SEP, and UASLP-FAI (Mexico); MBIE (New Zealand); PAEC (Pakistan); MSHE and NSC (Poland); FCT (Portugal); JINR (Dubna); MON, RosAtom, RAS and RFBR (Russia); MESTD (Serbia); SEIDI and CPAN (Spain); Swiss Funding Agencies (Switzerland); MST (Taipei); ThEPCenter, IPST, STAR and NSTDA (Thailand); TUBITAK and TAEK (Turkey); NASU and SFFR (Ukraine); STFC (United Kingdom); DOE and NSF (USA).

Individuals have received support from the Marie-Curie programme and the European Research Council and EPLANET (European Union); the Leventis Foundation; the A. P. Sloan Foundation; the Alexander von Humboldt Foundation; the Belgian Federal Science Policy Office; the Fonds pour la Formation \`a la Recherche dans l'Industrie et dans l'Agriculture (FRIA-Belgium); the Agentschap voor Innovatie door Wetenschap en Technologie (IWT-Belgium); the Ministry of Education, Youth and Sports (MEYS) of the Czech Republic; the Council of Science and Industrial Research, India; the HOMING PLUS programme of Foundation for Polish Science, cofinanced from European Union, Regional Development Fund; the Compagnia di San Paolo (Torino); the Thalis and Aristeia programmes cofinanced by EU-ESF and the Greek NSRF; and the National Priorities Research Program by Qatar National Research Fund.
\ifthenelse{\boolean{cms@external}}{\vspace*{12ex}}{}
\bibliography{auto_generated}

\cleardoublepage \appendix\section{The CMS Collaboration \label{app:collab}}\begin{sloppypar}\hyphenpenalty=5000\widowpenalty=500\clubpenalty=5000\textbf{Yerevan Physics Institute,  Yerevan,  Armenia}\\*[0pt]
V.~Khachatryan, A.M.~Sirunyan, A.~Tumasyan
\vskip\cmsinstskip
\textbf{Institut f\"{u}r Hochenergiephysik der OeAW,  Wien,  Austria}\\*[0pt]
W.~Adam, T.~Bergauer, M.~Dragicevic, J.~Er\"{o}, C.~Fabjan\cmsAuthorMark{1}, M.~Friedl, R.~Fr\"{u}hwirth\cmsAuthorMark{1}, V.M.~Ghete, C.~Hartl, N.~H\"{o}rmann, J.~Hrubec, M.~Jeitler\cmsAuthorMark{1}, W.~Kiesenhofer, V.~Kn\"{u}nz, M.~Krammer\cmsAuthorMark{1}, I.~Kr\"{a}tschmer, D.~Liko, I.~Mikulec, D.~Rabady\cmsAuthorMark{2}, B.~Rahbaran, H.~Rohringer, R.~Sch\"{o}fbeck, J.~Strauss, A.~Taurok, W.~Treberer-Treberspurg, W.~Waltenberger, C.-E.~Wulz\cmsAuthorMark{1}
% ...
\vskip\cmsinstskip
% ...
1:~~Also at Vienna University of Technology, Vienna, Austria\\
2:~~Also at CERN, European Organization for Nuclear Research, Geneva, Switzerland\\
% ...

\end{sloppypar}
\end{document}